\documentclass{aa} 
\usepackage{graphicx}
\usepackage{lscape}
\usepackage{longtable}
\usepackage{arydshln}
\usepackage{amsmath}
\usepackage{txfonts}
\usepackage{natbib}
\usepackage[htt]{hyphenat}
\usepackage{breqn}
\usepackage[normalem]{ulem}

\usepackage[dvipsnames]{xcolor}
\usepackage[colorlinks=true]{hyperref}
\hypersetup{linkcolor=red,citecolor=NavyBlue,filecolor=blue,urlcolor=magenta}

\bibpunct{(}{)}{;}{a}{}{,} 

\begin{document}

\title{AB Aur, a Rosetta stone for studies of planet formation (III): continuum observations at 2 and 7 mm.} 

\author{P. Rivi\`ere-Marichalar \inst{1}
\and E. Mac\'ias  \inst{2}
\and C. Baruteau  \inst{3}
\and A. Fuente    \inst{4}
\and R. Neri  \inst{5}
\and \'A. Ribas   \inst{6}
\and G. Esplugues \inst{1}
\and D. Navarro-Almaida \inst{7}
\and M. Osorio \inst{8}
\and G. Anglada \inst{8}
}

\institute{Observatorio Astron\'omico Nacional (OAN,IGN), Calle Alfonso XII, 3. 28014 Madrid, Spain 
                   \email{p.riviere@oan.es}           
	\and European Southern Observatory, Karl-Schwarzschild-Str. 2, 85748 Garching bei M\"unchen, Germany 
	\and IRAP, Universit\'e de Toulouse, CNRS, UPS, F-31400 Toulouse, France 
	\and Centro de Astrobiolog\'ia (CSIC/INTA), Ctra. de Torrej\'on a Ajalvir km 4, 28806 Torrej\'on de Ardoz, Spain 
        \and Institut de Radioastronomie Millim\'etrique, 300 rue de la Piscine, F-38406 Saint-Martin d'H\`eres, France 
	\and Institute of Astronomy, University of Cambridge, Madingley Road, Cambridge CB3 0HA, UK 
	\and Universit\'e Paris-Saclay, CEA, AIM, D\`epartement d'Astrophysique (DAp), F-91191 Gif-sur-Yvette, France 
        \and Instituto de Astrof\'isica de Andaluc\'ia (CSIC), Glorieta de la Astronom\'ia s/n, E-18008 Granada, Spain 
}

\authorrunning{Rivi\`ere-Marichalar et al.}
\titlerunning{AB Aur, a Rosetta stone for studies of planet formation (III)}
\date{}

 \abstract 
{Observational constraints on dust properties in protoplanetary disks are key to better understanding disks' evolution, their dynamics, and the pathway to planet formation, but also surface chemistry, the main driver of chemical complexity.}
{We continue our exploration of the protoplanetary disk around AB Aur by characterizing its dust properties at different millimeter wavelengths.}
{We present new ALMA observations at 2.2 mm and VLA observations at 6.8 mm. Together with previous ALMA and NOEMA observations at 0.87 and 1.1 mm, these new observations are used to compute global spectral index profiles as well as spectral index maps to probe the dust properties throughout the disk. On the interpretation side, we present the results of a simple isothermal slab model to help constrain dust properties along the non-axisymmetric ring of continuum emission outside the millimeter cavity. We also present new results of dust radiative transfer calculations based on a disc-planet hydrodynamical simulation to explain how the azimuthal contrast ratio of the ring emission varies with millimeter wavelength.}
{The spectral energy distribution and the radial profiles of the spectral index indicate that the radiation from the compact source towards the center is not dominated by dust thermal emission, but most likely by free-free emission originating in the radio jet: it constitutes 93\% of the emission at 6.8 mm, and 37\% at 0.87 mm. The protoplanetary disk has a typical spectral index of 2.3, computed using the 0.87, 1.1, and 2.2 mm bands. We estimate a dust disk mass of 8$\rm \times 10^{-5}$ M$\rm _{\odot}$ which, assuming a mean gas-to-dust ratio of 40, gives a total disk mass of 3.2$\rm \times 10^{-3}$ M$\rm _{\odot}$. The azimuthal contrast ratio of the ring outside the millimeter cavity is smaller at 2.2 mm than at 1.1 mm, in agreement with previous findings. The VLA image shows several knots of $5\sigma$ emission all along the ring, which, with the help of our dust radiative transfer calculations, are consistent with the ring emission being nearly axisymmetric at that wavelength. The decrease in the azimuthal contrast ratio of the ring emission from 0.87 to 6.8 mm can be explained by a dust-losing decaying vortex at the outer edge of a planet gap.}
{} 


\maketitle

\section{Introduction} 
Planets are born in the protoplanetary disks that surround young stars. Before the formation of planets, dust grains must grow in size. Understanding the evolution of dust is, therefore, crucial for theories of planet formation and evolution, and to explain the observed diversity of planetary systems. The evolution of dust is, in turn, affected by planet formation. Structures such as pressure maxima may efficiently trap dust particles, thus becoming pristine locations for further planet growth. 
This can be the case for instance at the outer edge of a planet gap, where a radial differentiation of dust sizes can occur as large grains get filtered while smaller grains can cross the gap \citep[e.g.,][]{Zhu2012}.  Other factors, such as the chemical composition of grains can alter the size distribution since average grain sizes are thought to be a factor of 100 smaller inside the water snow line \citep{Banzatti2015}. 

We can get insight into the dust properties by using mm and sub-mm continuum observations. By combining observations at different wavelengths we can compute spectral indices that inform us about the dust grain size distribution. Assuming optically thin continuum emission, the spectral index ($S_\nu \propto \nu^\alpha$) is related to the dust opacity ($\kappa_\nu \propto \nu^\beta$) in the Rayleigh-Jeans regime through $\alpha = \beta -2$. For the sub-micron-sized grains in the ISM, $\beta \sim$1.7 and becomes smaller as dust grains grow in size \citep{Miyake1993}, approaching zero in protoplanetary disks. Nowadays, thanks to interferometers such as ALMA, NOEMA, and the VLA, spectral index maps can be computed at a great angular resolution allowing us to study radial and azimuthal variations of the spectral index throughout the disk \citep{Perez2012, Perez2015, Pinilla2015, Casassus2015, Carrasco2019, Macias2019, Macias2021, Sierra2021, Ueda2022}. 

AB Aur is an HAe star located at $\approx$162.9 pc from the Sun \citep{GAIA2018} that is surrounded by a transition disk which has been studied in great detail. The source has been extensively observed at millimeter wavelengths using single-dish telescopes and interferometers. Such observations covered continuum emission as well as emission from molecular species. In terms of molecular emission, the list of species mapped towards the source includes HCO$\rm ^+$, HCN, $\rm ^{12}$CO, $\rm ^{13}$CO, SO,  H$\rm _2$CO, and H$\rm _2$S \citep{Riviere2019, Riviere2020, Riviere2022}, with radial emission peaks ranging from  0.9$\arcsec$ to 1.4$\arcsec$. Radial profiles show strong chemical segregation between molecular species, with differences as high as 100 au in the position of their peaks \citep{Riviere2020}.

Resolved observations have shown that the continuum emission extends out to $\sim$2.3$\arcsec$ \citep[$\rm \sim$370 au, ][]{Riviere2020} and features a large ($\sim$100 au-wide) cavity with an asymmetric ring of emission just outside the cavity \citep{Pietu2005, Tang2012, Fuente2017}. Oddly enough, the azimuthal contrast ratio of the flux along the ring decreases with increasing wavelength, which is opposite to what is expected with a dust trap \citep{Fuente2017}. The hydrodynamical simulations and dust radiative transfer calculations of \citet{Fuente2017} showed that this feature could be explained by a decaying vortex at the outer edge of a massive planet gap, with the vortex progressively losing the dust particles it had efficiently trapped before decaying. It can therefore be pictured as a decaying dust trap, although in the following we will simply refer to it as "dust trap". The simulations of \citet{Fuente2017} involved a $\rm \sim 2 ~M_{Jup}$ planet at 98 au from the central star and, interestingly, an embedded protoplanet at a similar distance (93 au) has been recently claimed by \cite{Currie2022} and \cite{Zhou2022} based on Subaru Telescope and Hubble Space Telescope observations. The twisted spiral seen in scattered light with SPHERE at about 30 au from the central star \citep{Boccaletti2020} is linked to this outer planet \citep{Currie2022}.

The aforementioned asymmetric ring is not the only continuum structure observed in the AB Aur disk. \cite{Tang2012} discovered a compact continuum source inside the cavity whose nature needs to be clarified. The HCO$\rm ^+$ map of the source depicts an outer disk with a decrease in intensity coincident with the dust cavity, a compact source toward the center, and a filamentary structure connecting the outer disk with the compact source in the center \citep{Riviere2019}. This filamentary structure was interpreted as the likely result of accretion at large scales. The compact emission in the center can be partially explained by the presence of a radio jet. AB Aur was first detected at cm wavelengths by \cite{Guedel1989}, who noted that the radio spectrum was not compatible with $\nu ^{0.6}$ as expected for free-free emission from a conical outflow \citep{Wright1975}. Using VLA observations in the mm and cm, \cite{Rodriguez2014} noted that the continuum emission at 7 mm was compatible with $\nu^{1.1}$ extrapolated from cm emission, thus consistent with free-free emission from a jet. Still, the exact nature of this compact source remains unknown. 

In this paper, we present continuum observations with ALMA, NOEMA, and VLA, and use them to compute spectral index maps. In Section \ref{Sect:obs_data_red} we describe the observations and explain the process of data reduction. In Section \ref{Sect:results} we describe our observational results, including spectral index maps and dust masses, and we present a dust model to help us interpret the multi-wavelength data. In Section \ref{Sect:interpretation} we compute dust masses, interpret our observations using a 1D isothermal slab model and present new dust radiative transfer calculations at 6.8 mm based on the disk-planet simulation results of \citet{Fuente2017}.
In Section \ref{Sect:discussion} we discuss the implications of our results. Finally, Section \ref{Sect:summary} summarizes our results.

\section{Observations and data reduction}\label{Sect:obs_data_red}

\begin{figure*}[t!]
\begin{center}
  \includegraphics[width=0.99\textwidth, trim=0mm 25mm 0mm 35mm, clip]{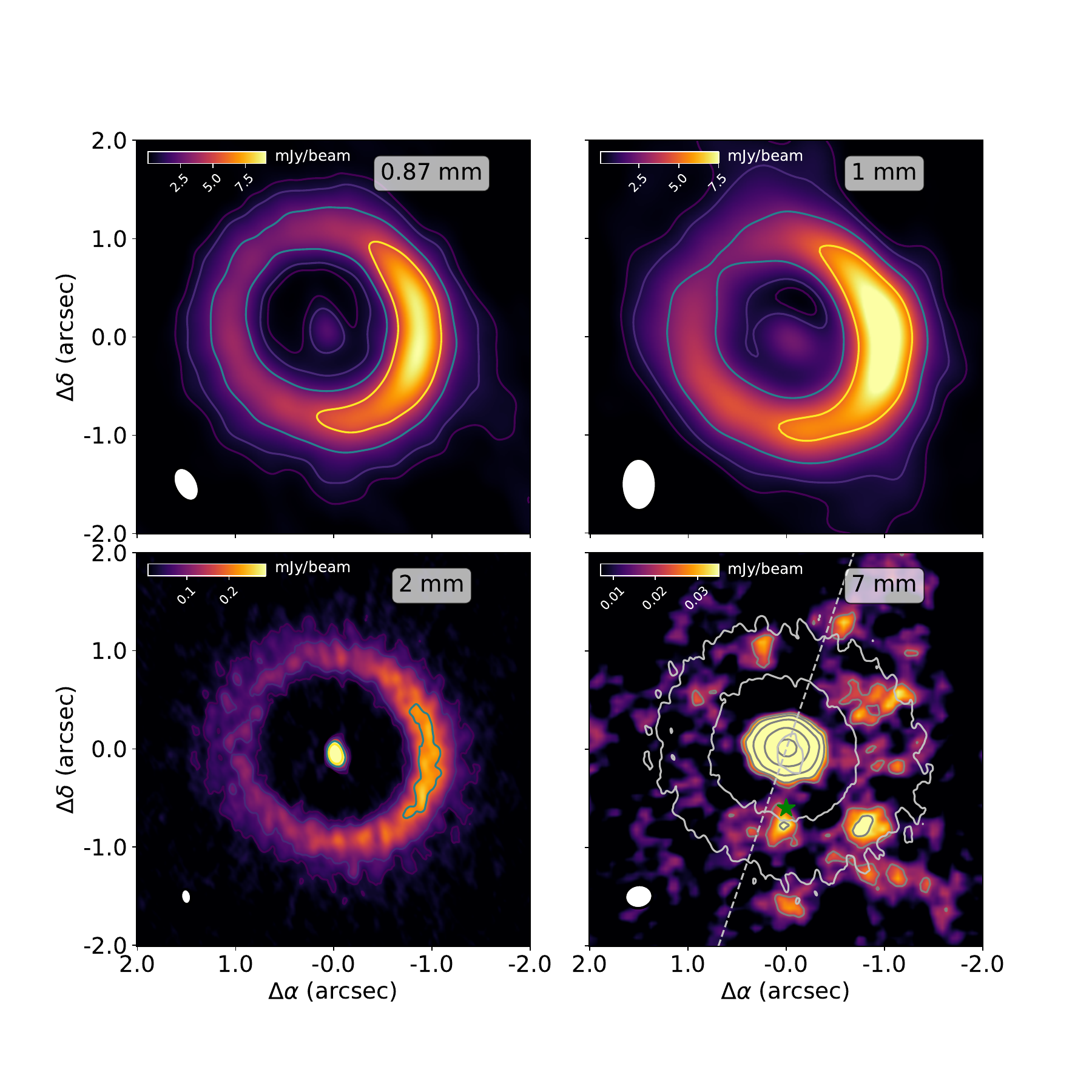}
  \caption{Top left: ALMA map at 0.87 mm. Top right: NOEMA map at 1.1 mm.  Bottom left: ALMA map at 2.2 mm. Bottom right: VLA map at 7 mm. The light grey contours in the bottom right panel depict 5$\sigma$ continuum emission at 2.2 mm from our ALMA map. The grey dashed line marks the position of the cm jet detected by \cite{Rodriguez2014}. The green star marks the position of the tentative protoplanet proposed by \cite{Currie2022}. The synthesized beam is shown in the bottom-left corner of each map. Contour levels in the  0.87, 1.1, and 6.8 mm maps are 5$\sigma$, 10$\sigma$, 25$\sigma$, and 50$\sigma$. The 2.2 mm map only shows 5$\sigma$, 10$\sigma$, and 25$\sigma$ levels. The map of the VLA emission at 7 mm also includes 3$\sigma$ contour levels. In all maps, the color bar indicates flux in mJy/beam.}
 \label{Fig:cont_maps}
\end{center}
\end{figure*}

\subsection{ALMA images}
We observed AB Aur at  2.2 mm (band 4) with ALMA using the 12m array (program code 2019.1.00579.S, PI: Asunci\'on Fuente) in two configurations. The most compact configuration was used on 2021 July 8 with 41 antennas covering baselines from 28 m to 3396 m. An RMS of 0.019 mJy/beam was reached with a total of 24 min of on-source observing time. The extended configuration was used on 2021 August 21 with 38 antennas covering baselines from 59 to 10803 m. An RMS of 0.010 mJy/beam was reached with a total of 83 min of on-source observing time. Both configurations used J0510+1800 as bandpass, pointing, and flux calibrator and J0512+2927 as phase calibrator. We observed four spectral windows at 137.791, 139.68, 149.791, and 151.749 GHz with a bandwidth of 2 GHz. The resulting central frequency is 144.764 GHz. We combined the visibilities from the two configurations using the \textit{tclean} method in CASA to produce high angular resolution maps. We used the multi-term multi-frequency synthesis deconvolver, the standard gridder, and natural weighting, resulting in a beam size of  $0\rlap.''16\times0\rlap.''10$.  The resulting map is shown in Fig. \ref{Fig:cont_maps}, bottom left panel. 

We retrieved archive ALMA Band 7 observations at 0.87 mm from the program 2012.1.00303.S (P.I. Tang, Y.-W.). The longest baseline employed was 1574.4 m. A total of 47.96 min of on-source observing time was used, reaching a sensitivity of 2$\rm \times 10^{-4}$ Jy/beam in the continuum image. The data were calibrated by the ALMA Archive and Data Retrieval (EU) team using J0510+1800 as flux calibrator and J0438+3004 as phase calibrator. Four spectral windows were observed, centered at 330.600 GHz, 332.012 GHz, 343.522 GHz, and 345.819 GHz. The resulting central frequency is 345.795 GHz. The visibilities were combined using CASA, and the map was built using tclean with a Hogbom deconvolver and natural weighting, resulting in a beam 0.362$\arcsec$ $\times$0.228$\arcsec$. The resulting map is shown in Fig. \ref{Fig:cont_maps}, top left panel. 

\subsection{NOEMA images}
We observed AB Aur with the NOEMA interferometer in its C configuration. The observations were carried out in January 2017 using Wideband Express (WIDEX, $\Delta \nu=2~MHz$). Continuum maps at 1.1 and 2.2 mm were presented in \cite{Fuente2017}, and the $\rm HCO^+$ and HCN 3-2 maps in \cite{Riviere2019}. We refer the reader to \cite{Fuente2017} for details about the data reduction process. In the present paper, we are only interested in the continuum emission at 1.1 mm, since our new ALMA map at 2.2 mm have a better resolution than the ones from  \cite{Fuente2017} (beam size $0\rlap."16$ $\times$ $0\rlap."10$ versus $0\rlap."71$ $\times$ $0\rlap."46$). The NOEMA image at 1.1 mm is shown in Fig. \ref{Fig:cont_maps}, top right panel.

\subsection{VLA images}
AB Aur was observed with the Very Large Array in the Q band in configurations C, B, and A. The A configuration observations (project code 15A-408) were taken on July 16 and August 03, 2015, with 16 minutes of on-source integration time each. The B configuration observations were carried out on June 22, 23, and 24 2019 (project code: 18B-132), and September 29, 2020 (project code: 20A-142). Each execution consisted of around 29 minutes of on-source integration time. Finally, the C configuration observations were taken on November 21, 2018 (project code: 18B-132), including 1 hour and 52 minutes of on-source integration time. The observations were taken using the maximum available bandwidth of 8 GHz, covering from 40 GHz to 48 GHz.

In all observations, 3C147 was used as flux calibrator, 3C84 as bandpass calibrator, and J0438+3004 as gain calibrator. The VLA calibration pipeline in CASA 5.4.2 (for observations taken earlier than 2020) and CASA 5.6.2 was used to perform the initial standard interferometric calibration. For each execution, the pipeline calibrated data were inspected and, in some cases, a few manual flags were applied to remove antennas with hardware issues.

Before combining all the executions, the data were aligned by correcting for the proper motions of AB Aur as reported by Gaia DR3 (Gaia collaboration et al. 2022). To correct for any possible misalignments caused by phase errors, we then performed phase self-calibration on the combined dataset. The mtmfs algorithm in tclean (CASA 6.5.1) was used during the self-calibration, with scale=0, nterms=2, and natural weighting. Given the limited signal-to-noise ratio of the data, we only performed two rounds of self-calibration; first combining scans, spectral windows, and correlations (i.e., gaintype = T), and in the second round combining only scans and spectral windows. This allowed us to increase the signal-to-noise ratio by 6\%. The imaging resulted in a beam of $0\rlap.''28\times0\rlap.''24$ , PA=-81 deg, and an rms of 7 $\mu$Jy/beam. The resulting map is shown in Fig. \ref{Fig:cont_maps}, bottom right panel.

\section{Results}\label{Sect:results}

\subsection{Continuum maps}\label{Sect:cont_maps}
In Fig. \ref{Fig:cont_maps} we show the NOEMA, ALMA, and VLA maps at 0.87, 1.1, 2.2, and 6.8 mm. The most prominent features are the asymmetric dust ring near 1\arcsec (hereafter, dust trap), the cavity, and the compact source toward the center. The extension of the dust trap changes from one band to another, as well as the precise position of the peak (albeit by a small margin). The 6.8 mm map shows 5$\sigma$ emission near the center and at a few azimuths along or around the ring. The compact source toward the center is observed in the four bands and is brighter than any position in the outer disk at 2.2 and 6.8 mm.

The 2.2 mm map (beam size $0\rlap."16$ $\times$ $0\rlap."10$)
spatially resolves the dust trap, which extends radially  from $0\rlap."7$ to $1\rlap."2$. It further allows us to separate the central source. To test whether the compact source in the center was resolved, we fitted a 2D Gaussian to 5$\sigma$ contour levels. We retrieved a value of 0$\rlap.''$164 for the major axis and of 0$\rlap.''$110 for the minor axis, versus 0$\rlap.''16\times0\rlap.''10$ for the beam. Considering the uncertainties, we cannot claim that the source is resolved with our 2 mm ALMA observations, and the synthesized beam size gives us an upper limit for its extent.
 
To compare the azimuthal contrast ratio along the ring in our maps, we convolved our ALMA data with the beam of our NOEMA map at 1.1 mm. Azimuthal contrast ratios were then computed as the ratio of the maximum to the minimum of the continuum flux along the ring in the maps convolved with the NOEMA beam. The azimuthal contrast ratios computed in this way are 3.1$\rm \pm $0.5 at 0.87 mm, 3.5$\rm \pm $0.2 at 1.1 mm, and 2.8$\rm \pm $0.6 at 2.2 mm. We could not compute the azimuthal contrast ratio of the 6.8 mm map due to the limited sensitivity of our VLA observations. If, instead of computing the azimuthal contrast ratio at the radial distance of the ring we do it with the radially-averaged disk intensity profiles, we obtain contrast ratios of 2.9$\rm \pm $0.7 at 0.87 mm, 3.1$\rm \pm$0.2 at 1.1 mm and 2.3$\rm \pm$ 0.7 at 2.2 mm, in agreement with the first method. Our results are in agreement with those derived in \cite{Fuente2017}. However, due to our large uncertainties, we cannot confirm that the contrast ratio is more pronounced at 1 mm than at 2 mm. 

We also computed the integrated continuum fluxes at the four frequencies by adding the fluxes inside 5$\sigma$ contours. The resulting fluxes are 204$\rm \pm$14 mJy at 0.87 mm, 108$\rm \pm$25 mJy at 1.1 mm, 23$\rm \pm$6 mJy at 2.2 mm, and 2.2$\rm \pm$1.1 mJy at 6.8 mm (see Table \ref{Tab:SED}). Furthermore, since our observations resolve the emission towards the center, we can separate that component from the outer disk and compute the fluxes from each region separately. These fluxes are shown in Table \ref{Tab:disk_masses}.

\begin{figure}[t!]
\begin{center}
  \includegraphics[width=0.5\textwidth]{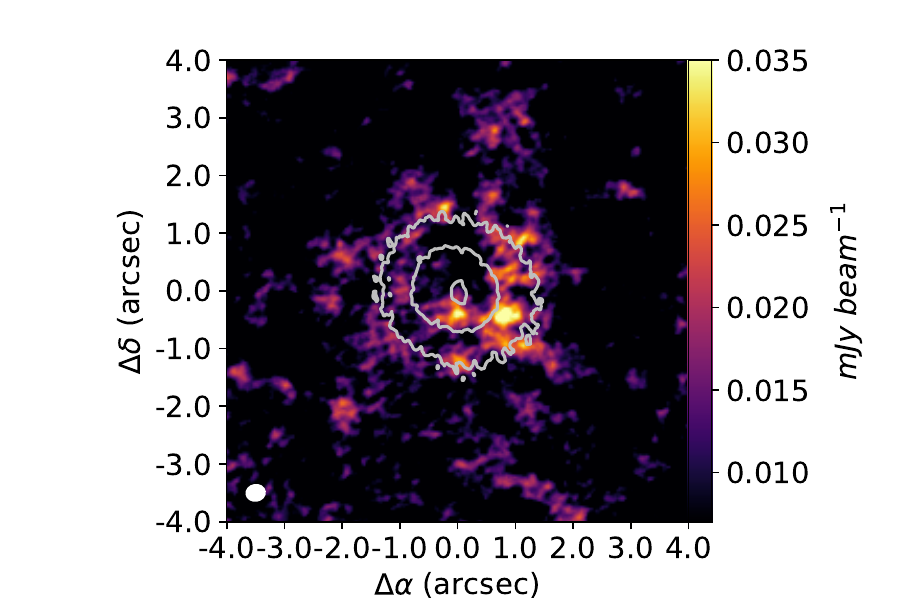}
  \caption{ VLA map at 6.8 mm after subtracting the compact source in the center from the visibilities.}
 \label{Fig:VLA_ring}
\end{center}
\end{figure}

The VLA map at 6.8 mm (beam size $0\rlap.''282\times0\rlap.''237$) is displayed in the bottom right panel of Fig.~\ref{Fig:cont_maps}.  ALMA continuum data at 2.2 mm are overlaid as contours to facilitate the comparison between both maps. Continuum emission at 6.8 mm is marginally detected at a distance of 1$\arcsec$ from the center. We detect 5$\sigma$ emission in at different azimuths, but the emission is more prominent to the south and west of the disk, around the position of the dust trap as well as toward the center. The compact source towards the center is not resolved. 

To determine the distribution of the weak continuum emission around the central source, an initial step involved fitting it directly in the uv-plane using a Gaussian profile. The source, which appeared in the 6.8-mm continuum as unresolved ($<$ 0.03\arcsec) and with a flux of 429 $\rm \pm$ 10 $\mu$Jy, was then directly subtracted from the visibilities to obtain an imprint of the underlying distribution of the outer continuum emission. To achieve maximum sensitivity, the visibilities were converted using their natural weights into a dirty map with a resolution of 0.47\arcsec x 0.43\arcsec and a PA of 80$^{\circ}$ degrees (Fig. \ref{Fig:VLA_ring}). Despite the relatively weak continuum emission, the resulting map shows that the emission is roughly consistent with being ring-shaped. Cleaning of the map was not attempted due to the very limited SNR, but an extrapolation of the visibilities to zero-spacing indicates that the flux in the outer structure is approximately 0.4 +/- 0.1 mJy (see Fig. \ref{fig:model6v8_uvplane}).

\subsection{Radial profiles}
\begin{figure}[t!]
\begin{center}
 \includegraphics[width=0.5\textwidth,trim = 0mm 0mm 0mm 0mm,clip]{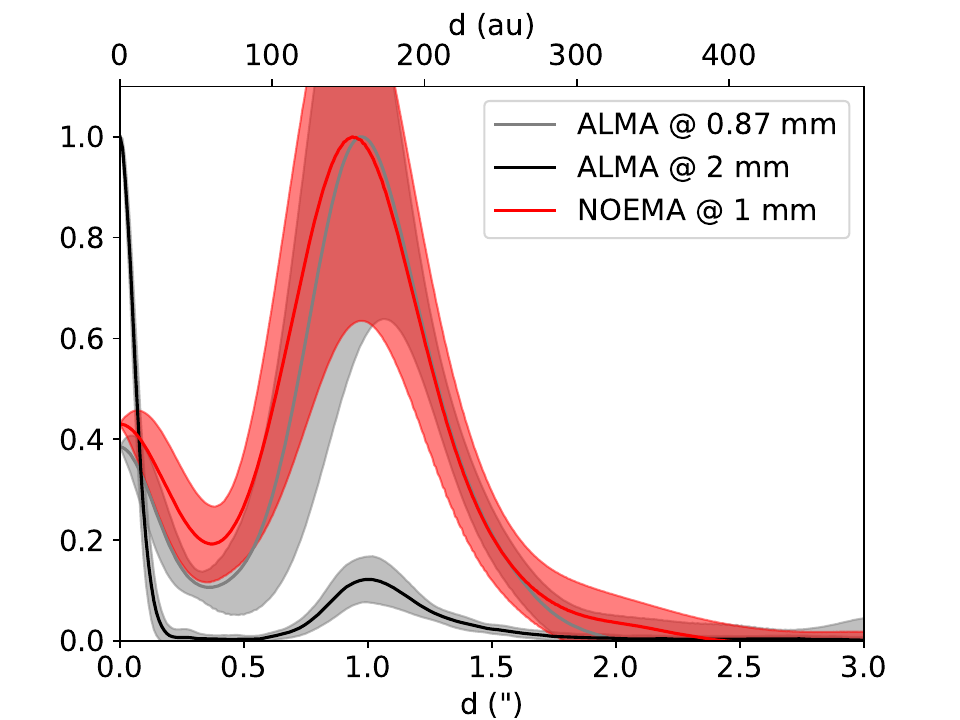}\\
  \caption{Azimuthally-averaged, normalized radial profiles of the 0.87mm, 1.1mm, and 2.2 mm continuum emission (solid curves, maps were deprojected with PA=-37$\rm ^{\circ}$ and i=24.9$\rm ^{\circ}$). Shaded areas show the azimuthal variations of the emission maps at each deprojected radius.}
 \label{Fig:radial_profile}
\end{center}
\end{figure}

Figure \ref{Fig:radial_profile} shows azimuthally averaged radial profiles of our deprojected continuum maps at 0.87, 1.1, and 2.2 mm. The maps were deprojected assuming $PA=-37^{\circ}$ and $i=24.9^{\circ}$ \citep{Riviere2020}. Due to the lack of emission at most azimuths, the radial profile of the 6.8 mm map is not shown. The 0.87 and 1.1 mm profiles peak slightly below 1.0$\arcsec$, that is at the deprojected radius of the continuum ring. In contrast, the profile at 2.2 mm peaks at the center as the emission from the compact source toward the star largely exceeds that of the continuum ring.

\subsection{Azimuthal profiles}
\begin{figure}[t!]
\begin{center}
 \includegraphics[width=1.08\hsize,trim = 0mm 0mm 0mm 0mm,clip]{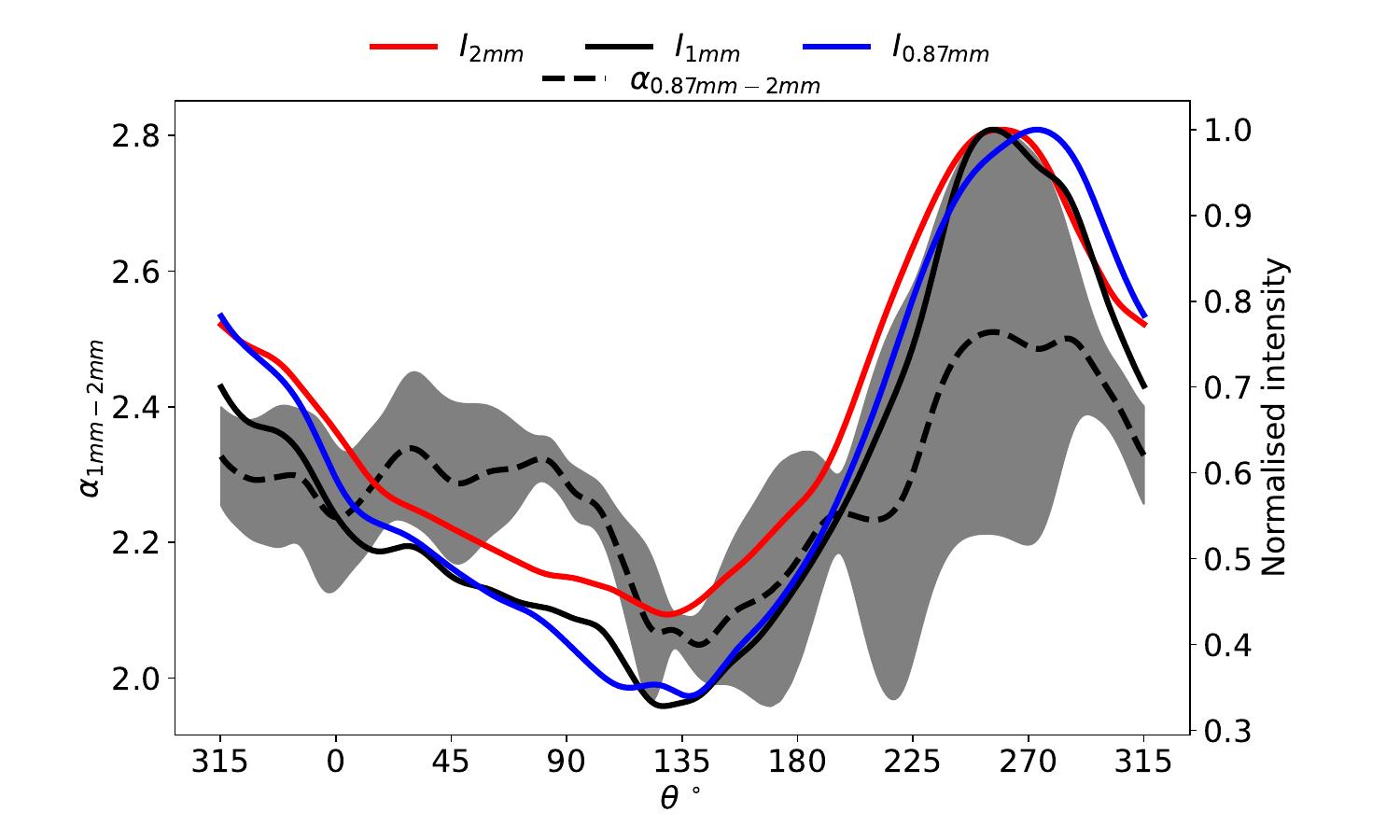}\\
  \caption{Radially averaged azimuthal profiles of the NOEMA and ALMA maps at 0.87mm (solid blue curve), 1.1 mm (solid black curve), and 2.2 mm (solid red curve) and of the $\alpha_{0.87{\rm mm}-1{\rm mm}-2{\rm mm}}$ spectral index (dashed curve). To compute the azimuthal profile the maps were deprojected with PA=-37$\rm ^{\circ}$ and i=24.9$\rm ^{\circ}$. The 0.87 and 2.2 mm maps were convolved with the beam of the 1 mm map for comparison purposes. }
 \label{Fig:az_profile}
\end{center}
\end{figure}

We show in Fig. \ref{Fig:az_profile} the radially averaged azimuthal profiles of our ALMA and NOEMA continuum maps at 0.87, 1.1, and 2.2 mm. The profiles were computed by averaging the intensity at each azimuth between $r=50$ au and $r=140$ au. The profile for the VLA map at 6.8 mm is not shown, since we lack emission at most azimuths. The radial averages were computed only along the disk. Furthermore, to allow for visual comparison and before computing the azimuthal profiles, the maps at 0.87 and 2.2 mm were convolved with the beam of the 1.1 mm map. The three profiles look very similar, but they peak at slightly different azimuthal angles. A more detailed comparison of the profiles requires observations with better angular resolution at 1.1 mm.

\subsection{Spectral energy distribution at mm wavelengths}\label{Sect:SED}
\begin{figure}[ht!]
\begin{center}
  \includegraphics[width=0.99\hsize]{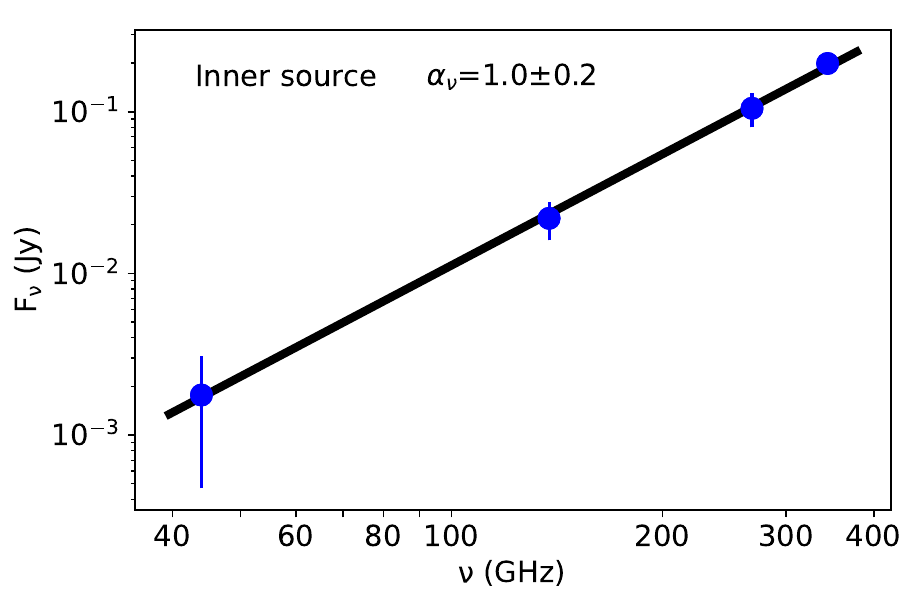}\\
  \includegraphics[width=0.99\hsize]{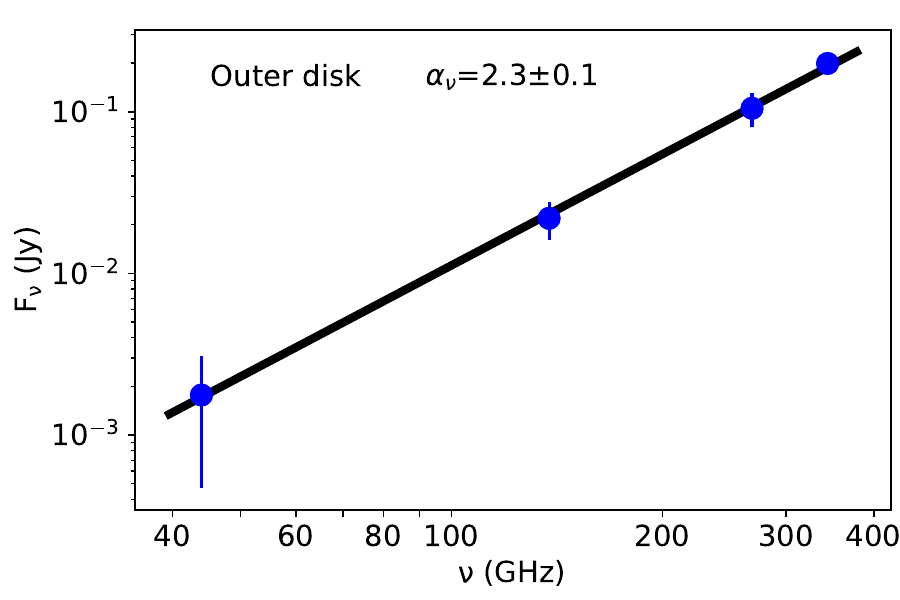}\\
  \caption{Spectral energy distribution at mm wavelengths. Top: for the inner source. Bottom: for the outer disk. The blue dots represent our observations at different wavelengths, while the black line is a linear fit to the data.}
 \label{Fig:SED}
\end{center}
\end{figure}

We used observations of AB Aur to build a spectral energy distribution (SED) at millimeter wavelengths. Since (sub)mm observations reported in this work separate the inner source from the outer disk, we computed the SED for each region separately. We first integrated the flux inside a 5$\sigma$ contour around the center to retrieve the flux from the inner compact source. This flux was later subtracted from the total flux to get the flux of the outer disk. Table \ref{Tab:disk_masses} summarizes the continuum fluxes employed. The resulting SEDs are shown in Fig. \ref{Fig:SED}. We measure a spectral index of 1.1$\rm \pm$0.1 for the compact inner source, and 2.3$\rm \pm$0.1 for the outer disk. 

\begin{table*}[ht!]
\caption{Continuum flux at mm wavelengths for the central source and the outer disk.}
\begin{center}
\begin{tabular}{ccccc}
\hline \hline
Wavelength & Flux outer disk & $\rm M_{dust}$ outer disk& Flux inner disk  &  $\rm M_{dust}$ inner disk \\
(mm) & (mJy) & ($\rm M_{Earth}$) &(mJy) & ($\rm M_{Earth}$) \\
\hline
0.87 & 200$\rm \pm$14 & 28$\rm \pm$2 & 3.3$\rm \pm$0.5 & 0.47$\rm \pm$0.07  \\
1.1 & 105$\rm \pm$25 & 26$\rm \pm$6  & 3.1$\rm \pm$0.6 & 0.46$\rm \pm$0.06 \\
2.2 &  21$\rm \pm$6 & 23$\rm \pm$6 & 1.3 $\rm \pm$ 0.3 & 0.50$\rm \pm$0.06\\
6.8 & 1.8$\rm \pm$1.3 & 23$\rm \pm$17 & 0.43$\rm \pm$0.09 & 0.35$\rm \pm$0.17\\
\hline
\end{tabular}
\end{center}
\label{default}
\label{Tab:disk_masses}
\end{table*}%

\subsection{Spectral index maps}\label{Sect:sp_index}
Since some of our maps resolve the inner and outer disks, we can further compute spectral index maps for pairs of wavelengths to study the spatial distribution of these spectral indices. 

We produced broadband spectral index maps by combining our maps at 0.87, 1.1, and 2.2 mm. Again, the map at 6.8 mm is left out of this exercise due to the lack of emission at most azimuths. To compute each spectral index map, we convolved the map with the highest angular resolution to reach the resolution of the one with the lowest angular resolution. Thus, we convolved the 0.87 mm map to the beam size of the 1 mm map to compute the $\alpha_{0.87{\rm mm}-1{\rm mm}}$ spectral index map, we convolved the 2 mm map to the beam size of the 0.87 mm map to compute the $\alpha_{0.87{\rm mm}-2{\rm mm}}$, and finally, we convolved the 2 mm map to the beam size of the 1 mm map to compute the $\alpha_{1{\rm mm}-2{\rm mm}}$. The resulting maps are shown in Fig.~\ref{Fig:sp_index_maps}. We also computed the spectral index map resulting from fitting 0.87, 1, and 2 mm continuum fluxes together (see Fig. \ref{Fig:sp_index_map_all}). In Table \ref{Tab:sp_index_inner_outer}, we present a summary of the mean spectral indices at different wavelength ranges and in the different regions of the disc. Pixels with s/n$<$3 in any of the bands employed in each map were masked out to compute these statistics. The compact source towards the center shows a lower spectral index, with mean values of 0.8 to 1.9. In contrast, the outer disc shows values in the range [2.2-2.8].

The trends with radius are easy to observe in the azimuthally averaged radial profiles of the spectral index maps shown in the lower panel in Fig. \ref{Fig:sp_index_maps}. The profiles show a trend for increasing spectral index with increasing radius until r$\sim$0.5-1.0$\arcsec$ beyond which it plateaus. We note that $\alpha_{0.87{\rm mm}-2{\rm mm}}$ shows a local maximum around r$\sim$0.5$\arcsec$, which could indicate sub-micron-sized grains inside the cavity and in the innermost parts of the protoplanetary disc. However, one should be aware that it could also be due to the beam extension.

In Fig. \ref{Fig:az_profile} we show the radially averaged azimuthal profile of the $\alpha_{0.87{\rm mm}-1{\rm mm}-2{\rm mm}}$ spectral index together along with the radially averaged azimuthal profiles of the continuum emission at 0.87, 1, and 2 mm. The continuum and the spectral index profiles peak at slightly different angular positions but, overall, the shape of the spectral index profile shows a strong similarity with that of the continuum emission.

 For the 6.8 mm continuum map, we can compute the spectral indices where 5$\sigma$ emission is detected. By doing so we retrieve 1.0$\rm \pm$0.1, 1.2$\rm \pm$0.2, and 0.9$\rm \pm$0.3 for $\alpha_{0.87{\rm mm}-6.8{\rm mm}}$, $\alpha_{1.1{\rm mm}-6.8{\rm mm}}$, and $\alpha_{2.2{\rm mm}-6.8{\rm mm}}$ for the compact source toward the center and 2.5$\rm \pm$0.2, 2.5$\rm \pm$0.2, and 2.4$\rm \pm$0.4 for the outer disk. We cannot compute the spectral index for the 7 mm map at all azimuths, but when the emission is detected we measure values from 2 to 2.8 along the ring.

Taking advantage of the frequency coverage of the NOEMA and ALMA observations, we produced a spectral index map computed by fitting a power law to individual pixels. Before SED fitting per pixel, we convolved all the maps to a common resolution given by the 1.1 mm map ($0\rlap.''53\times0\rlap.''35$). The resulting map is shown in Fig. \ref{Fig:sp_index_map_all}, as well as the fitted SED at different positions in the disk. We also include in Fig.  \ref{Fig:sp_index_map_all} the radial profile of the spectral index map (panel b).  Again we note the sub-thermal emission toward the center, an increase in the spectral index inside the cavity, and a more or less flat spectral index along the disk. The mean spectral index for the compact source toward the center is 1.1$\rm \pm$0.3, and for the disc, it is 2.3$\rm \pm$0.2.

\begin{table}[htbp]
\caption{Spectral indices in different wavelength ranges and disk regions.}
\begin{center}
\begin{tabular}{ccc}
\hline \hline
Wavelength range & $\alpha_{\rm inner}$ & $\alpha_{\rm outer}$ \\
(mm) & -- & --   \\
\hline
0.87-1.1 & 1.3$\rm \pm$0.7  & 2.8$\rm \pm$0.7  \\
0.87-2.2 & 0.8$\rm \pm$0.3 & 2.4$\rm \pm$0.2\\
1.1-2.2 & 1.6$\rm \pm$0.7 & 2.2$\rm \pm$0.2 \\
0.87-1.1-2.2 & 1.1$\rm \pm$0.2 & 2.3$\rm \pm$0.4 \\
\hline
\end{tabular}
\end{center}
\label{default}
\label{Tab:sp_index_inner_outer}
\end{table}%

\begin{table}[ht]
\caption{Continuum fluxes at sub-mm, mm, and cm wavelengths.}
\begin{center}
\begin{tabular}{lll}
\hline \hline
Wavelength & Flux & Ref \\
(mm) & (mJy) & -- \\
\hline
0.45 & 2200 $\rm \pm$ 100 & \cite{Sandell2011} \\
0.85 & 350  $\rm \pm$ 20 & \cite{Sandell2011} \\ 
0.87 & 204 $\rm \pm$ 14 & This work \\
1.1 & 108 $\rm \pm$ 25 & \cite{Fuente2017} \\
1.3 & 92 $\rm \pm$ 2 &  \cite{Riviere2020}  \\
1.88 & 42 $\rm \pm$ 9 & \cite{Riviere2022} \\
2.2 & 23.2 $\rm \pm$ 5.9 & This work \\
6.8 & 2.2 $\rm \pm$ 1.3 & This work \\
33 & 0.136 $\rm \pm$ 0.006 & \cite{Rodriguez2014} \\
35 & 0.146 $\rm \pm$ 0.025 & \cite{Guedel1989} \\
35 & 0.112 $\rm \pm$ 0.025 & \cite{Guedel1989} \\
35 & 0.140 $\rm \pm$ 0.024 & \cite{Skinner1993}\\
35 & 0.150 $\rm \pm$ 0.020 & \cite{Skinner1993}\\
35 & 0.200 $\rm \pm$ 0.050 & \cite{Rodriguez2007}\\
40 & 0.167 $\rm \pm$ 0.030 & \cite{Dzib2015}\\
60 & 0.119 $\rm \pm$ 0.020 & \cite{Guedel1989} \\
67 & 0.089 $\rm \pm$ 0.019 & \cite{Dzib2015}\\
\hline
\end{tabular}
\end{center}
\label{default}
\label{Tab:SED}
\end{table}%

\section{Interpretation}\label{Sect:interpretation}

This section is dedicated to the interpretation of our observational data. We start in Section~\ref{sec:dustmass} with estimates of the dust mass along the ring of continuum emission and in the entire disk. Section~\ref{sec:slabmodel} continues with a simple 1D isothermal slab model aimed at getting some insight into the dust properties (maximum size, surface density) along the ring. We then present in Section~\ref{sec:RTcalcs} synthetic maps of continuum emission at 1.1, 2.2, and 6.8 mm obtained from dust radiative transfer calculations that are based on the hydrodynamical model of \citet{Fuente2017}.

\subsection{Dust mass}
\label{sec:dustmass}
Assuming that the continuum emission at millimeter wavelengths is optically thin, we can use the measured millimeter fluxes to derive the dust mass using:

\begin{equation}\label{Eq:Mdust}
M_d = \frac{F_\nu d^2}{\kappa_\nu B_\nu (T_{\rm d})}
\end{equation}
\citep{Beckwith1990}, where $F_\nu$ is the averaged intensity at the observed frequency $\nu$, $d$ is the distance to the source, $\kappa_\nu$ the dust's absorption opacity at the observed frequency $\nu$, $B_\nu$ the black body intensity at temperature $T_{\rm d}$ and frequency $\nu$, and $T_{\rm d}$ the dust temperature. Flux densities used are listed in Table \ref{Tab:SED}. We adopt the standard prescription for the dust's absorption opacity \citep{Andrews2005}:

\begin{equation}\label{Eq:Kappa}
\kappa_\nu = \kappa_{0} \left( \frac{\nu}{\nu_0} \right)^\beta
\end{equation}

\noindent with $\nu_0$ = 230 GHz and $\kappa_0 = \kappa(\nu_0)= 2.3$ cm$^2$ g$^{-1}$. The $\beta$ value can be derived from the spectral index $\alpha$ at millimeter wavelengths obtained in Sec. \ref{Sect:SED} via $\beta = \alpha -2$. 
Using the four bands available, we get $\beta$=0.2$\rm \pm$0.1. For the dust temperature ($T_{\rm dust}$) we assume a value of 39~K based on previous work \citep{Riviere2020}. We stress that the value of $\kappa_0$ in Eq.~(\ref{Eq:Kappa}) is subject to large uncertainties related to the dust's size distribution, composition, and porosity. The computed dust masses, which are listed in Table \ref{Tab:disk_masses}, thus have to be considered with caution.

The resulting dust masses range from 6.6$\rm \times 10^{-5}~M_{\odot}$ to 9.6$\rm \times 10^{-5}~M_{\odot}$, with an average value of (8.0$\rm \pm$1.0)$\rm \times 10^{-5}~M_{\odot}$. The dust mass of the ring alone is computed assuming $\alpha$=2.3$\rm \pm$0.4 and by taking the flux of the outer disk alone (see Table \ref{Tab:disk_masses}). The mean mass derived is (7.5$\rm \pm$0.6)$\rm \times 10^{-5}~M_{\odot}$. We note that \cite{Fuente2017} predicted an optical depth $\sim$0.9 at 1.1 mm at the peak of the dust trap, and thus our estimates are lower limits to the actual disk mass. However, since this optical depth is only reached at the dust trap emission peak, the impact on the real disk mass should be minor.

\begin{figure*}[t!]
\begin{center}
 \includegraphics[width=1\textwidth]{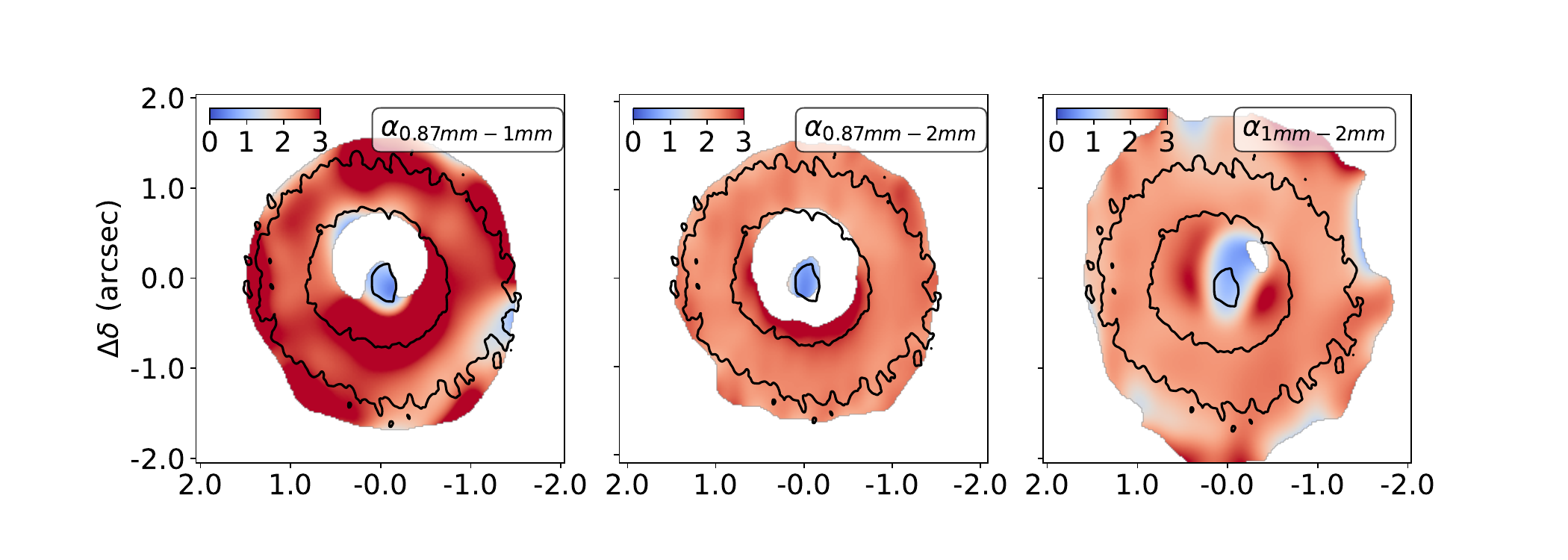}\\
 \includegraphics[width=1.0\textwidth, trim=0mm 0mm 0mm 0mm, clip]{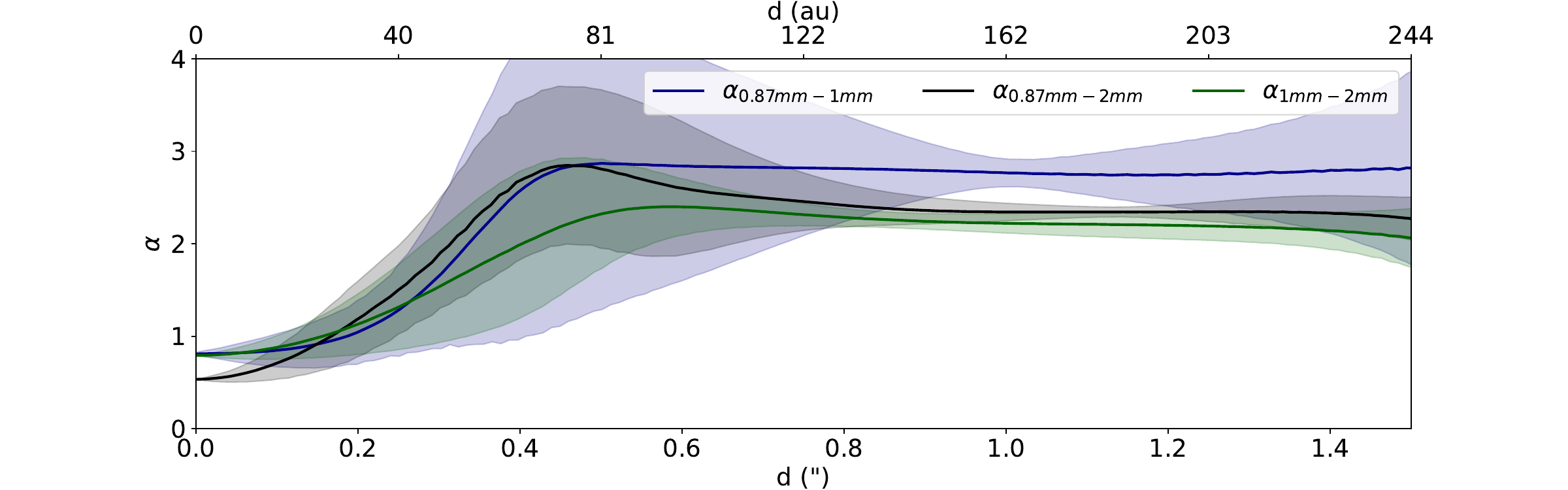}\\
  \caption{Top: Spectral index maps between 0.87 mm and 1.1 mm (left), 0.87 mm and 2 mm (middle), and 1.1 and 2.2 mm (right). Black contours depict 5$\sigma$ continuum emission at 2.2 mm. Bottom: Azimuthally averaged radial profiles of the spectral index maps between 0.87 and 1.1 mm, 0.87 and 2.2 mm, and 1.1 and 2.2 mm. The shaded areas represent the uncertainty in the radial profiles.}
 \label{Fig:sp_index_maps}
\end{center}
\end{figure*}

\begin{figure*}[t!]
\begin{center}
 \includegraphics[width=1.2\textwidth, trim=90mm 0mm 0mm 0mm, clip]{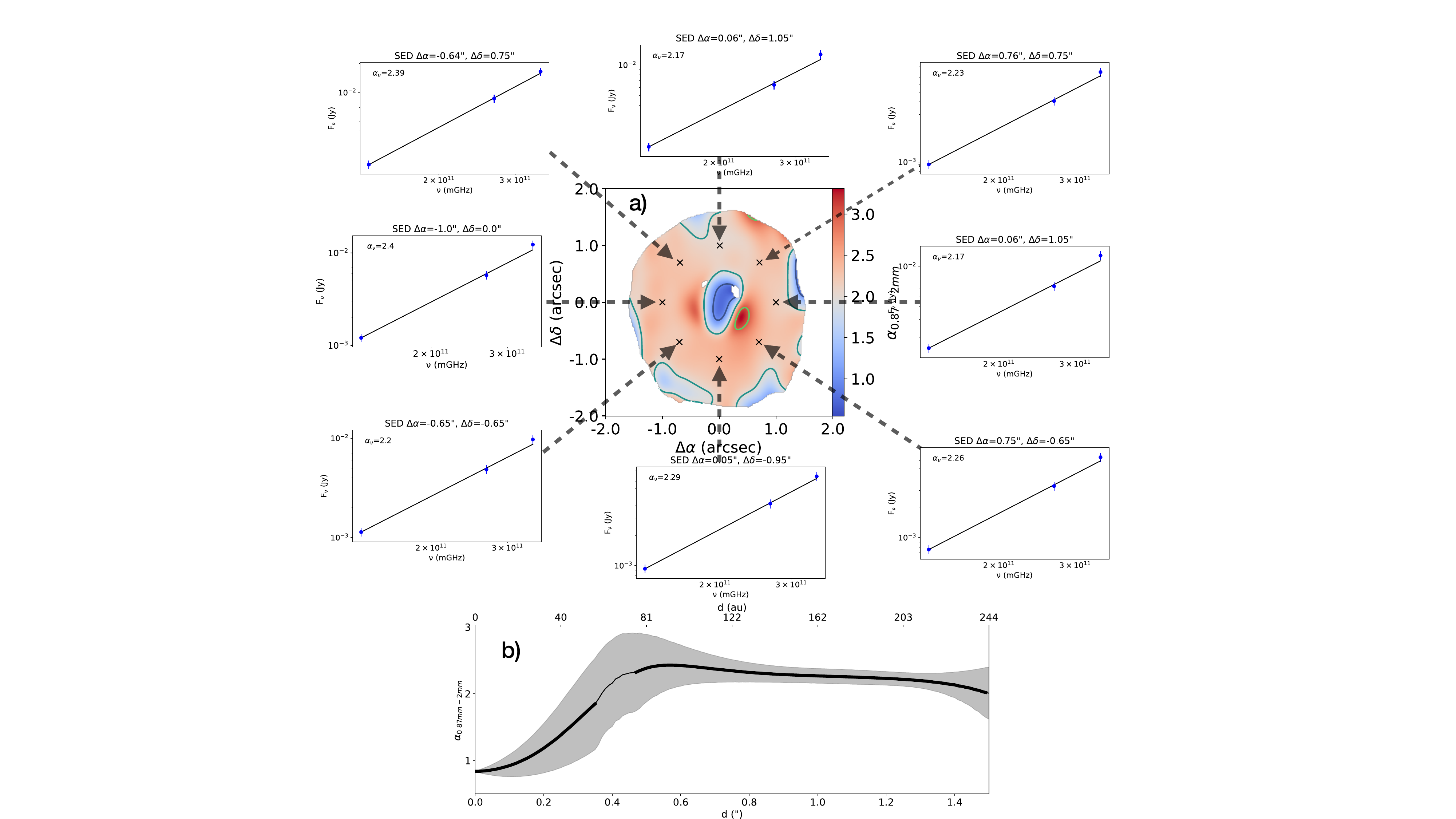}\\
  \caption{a) Spectral index map computed using the 0.87 mm, 1.1 mm, and 2.2 mm data. Also shown are the SED at different positions in the disk. b) Radial profile of the spectral index map computed using the 0.87 mm, 1.1 mm, and 2.2 mm data.}
 \label{Fig:sp_index_map_all}
\end{center}
\end{figure*}

\subsection{A 1D isothermal dust slab model}
\label{sec:slabmodel}
Following the approach in \cite{Macias2021}, we used the multiwavelength ALMA and NOEMA observations of AB Aur to characterize the properties of the dust around AB Aur, namely the density, dust temperature, and dust particle size distribution \citep{Perez2015, Carrasco2019, Macias2019}. However, given the large cavity present in AB Aur, we are more interested in modeling the radially averaged azimuthal profile to compare the distribution of grain sizes between the crescent and the rest of the ring. To that goal, we averaged the disk emission at all azimuths between r=$0\rlap."7$ and r=$1\rlap."25$. 

\subsubsection{Assumptions}
The model computes the intensity at a given azimuth as a 1D isothermal slab \citep{Carrasco2019} described by equation

\begin{equation}
    I_{\nu} = B_{\nu}(T_{\rm d}) [(1-\textrm{exp}(-\tau_{\nu}/\mu))+\omega_{\nu}F(\tau_{\nu},\omega_{\nu})],
\end{equation}
where, again, $T_{\rm d}$ is the dust temperature, $B_{\nu}(T_{\rm d})$ is the black body emission at the dust temperature ($T_{\rm d}$) and at frequency $\nu$, $\tau_{\nu}=\Sigma_{\rm d} \chi_{\nu}$ is the optical depth, $\Sigma_{\rm d}$ is the dust surface density, $\chi_{\nu}$ is the dust opacity including absorption and scattering, $\omega_{\nu}=\sigma_{\nu}/(\kappa_{\nu}+\sigma_{\nu})$ is the dust albedo, and $\mu=\cos(i)$, where $i$ is the inclination angle of the disk, which is 24.9$^{\circ}$ in the case of AB Aur. The function $F$ is defined by 
\begin{equation}
\begin{gathered}
    F(\tau_{\nu},\omega_{\nu})=
    \frac{1}{\textrm{exp}(-\sqrt{3}\epsilon_{\nu}\tau_{\nu})(\epsilon_{\nu}-1)-(\epsilon_{\nu}+1)} \times\\
    \left[\frac{1-\textrm{exp}(-(\sqrt{3}\epsilon_{\nu}+1/\mu)\tau_{\nu})}{\sqrt{3}\epsilon_{\nu}\mu+1} +
    \frac{\textrm{exp}(-\tau_{\nu}/\mu)-\textrm{exp}(-\sqrt{3}\epsilon_{\nu}\tau_{\nu})}{\sqrt{3}\epsilon_{\nu}\mu-1}\right] 
\end{gathered}
\end{equation}
and  $\epsilon_{\nu}=\sqrt{1-\omega_{\nu}}$.

Assuming a fixed dust composition as well as a dust particle size distribution ($n(a)\propto a^{-p}$, where $a$ is the dust grain radius) the dust absorption and scattering laws can be computed varying the minimum and maximum particle size ($a_{\rm min}$, $a_{\rm max}$) and the slope of the particle size distribution ($p$). However, as noted in \cite{Macias2021} the effect of $a_{\rm min}$ in the (sub)millimeter opacity is negligible as long as it is small, and thus we fixed  $a_{\rm min}$=$0.01$ $\mu$m in our model. For $p$ we assume the standard ISM value of 3.5 \citep{Mathis1977}. Thus, the free parameter regarding dust size in our models is $a_{\rm max}$. Following \cite{Macias2019} we use the dust composition assumed by the DSHARP Large Programme \citep{Birnstiel2018}. Given our assumptions, the intensity at a given azimuth and a given wavelength can be computed as a function of the dust temperature $T_{\rm d}$, the dust surface density $\Sigma_{\rm d}$ and the maximum grain size $a_{\rm max}$.

To retrieve the model parameters we followed a Bayesian approach, and performed a Monte-Carlo Markov Chain (MCMC) exploration of the parameter space.  We use a log-normal likelihood function:
\begin{equation}
    \textrm{ln}P(\bar{I}(\theta)|\Theta)=-0.5\sum_i \left( \left( \frac{\bar{I_i}-I_{m,i}}{\hat{\sigma}_{\bar{I},i}} \right)^2 + \textrm{ln}(2\pi\hat{\sigma}_{\bar{I},i}^2)\right) ,
\end{equation}
with $\Theta$ the vector of model parameters ($T_{\rm d}$, $\Sigma_{\rm d}$, and $a_{\rm max}$), $\bar{I_i}$ is the radially averaged intensity at azimuth $\theta$ and frequency $\nu_i$, $I_{m,i}$ is the model intensity at this same azimuth and frequency, and $\hat{\sigma}_{\bar{I},i}$ is the uncertainty at azimuth $\theta$, $\hat{\sigma}_{\bar{I},i} = \sqrt{\sigma_{\bar{I},i}^2 + (\delta_i \bar{I}_i)^2}$, where $\sigma_{\bar{I},i}$ is the error of the mean obtained from the radially averaged intensity profiles and $\delta_i$ is the absolute flux calibration error at each frequency. We assumed a 10\% error at all bands. Finally, we assumed a Gaussian prior for $T_{\rm d}$ with a mean value of 39 K and $\sigma$=4, and flat, non-informative priors for $\Sigma_{\rm d}$ (10$\rm ^{-4}<\Sigma_{\rm d}/g ~cm^{-2} <$10) and $a_{\rm max}$ (10$\rm ^{-2} < a_{\rm max}/cm<$10).

Given the limited sensitivity of our 6.8 mm VLA maps, we performed the model fitting using only the 0.87, 1.1, and 2.2 mm bands.  To allow for the comparison of the azimuthal profiles, we convolved them to a common beam, that of the NOEMA map at 1.1 mm ($0\rlap.''53\times0\rlap.''35$, PA=0.7$^{\circ}$). 

\subsubsection{Results}
We show in Fig. \ref{Fig:dust_RT_model_results} the posterior distribution resulting from our MCMC exploration of the parameter space. The shaded regions are limited by the 16\% and 84\% confidence levels. We discuss first the posterior distribution of the dust temperature. As can be seen in Fig. \ref{Fig:dust_RT_model_results}, the posterior strongly follows the prior for the dust temperature, and thus the model fitting does not add new information regarding this parameter. The other two free parameters  (the surface density and the maximum grain size) however, show some modulation along the azimuthal direction. In both cases, the variations are within the computed uncertainties, and thus the following trends are only tentative and require observations at more bands and with better sensitivity to be confirmed.

The dust surface mass density follows the continuum emission and may peak at the position of the dust trap, delimited in Fig. \ref{Fig:dust_RT_model_results} by the two red dashed vertical lines. The dust surface density reaches a maximum of  (7.7$\rm \pm 1.4)\times 10^{-3} g ~cm^{-2}$ at $\theta$=267$^{\circ}$, while the 0.87, 1.1, and 2.2 mm continuum maps peak at $\theta$=252$^{\circ}$, 252$^{\circ}$ and 245$^{\circ}$, respectively. Furthermore, it reaches a minimum of  (2.9$\rm \pm 1.4)\times 10^{-3} g ~cm^{-2}$ at $\theta$=267$^{\circ}$  at $\theta$=137$^{\circ}$, while the continuum maps at 0.87, 1.1, and 2.2 mm reach a minimum at $\theta$=138$^{\circ}$, 130$^{\circ}$ and 126$^{\circ}$, respectively. Thus, as expected, the dust surface density azimuthal profile  may follow the continuum mm brightness azimuthal profile. The azimuthal contrast in the surface density profile is 2.6, similar to the contrast measured in the continuum maps (2.9, 3.1, and 2.3 at 0.87,1.1 and 2.2 mm respectively, with a mean value of 2.8, see Sect.~\ref{Sect:cont_maps}). As mentioned in the previous paragraph, these results must be considered with caution due to the large uncertainty in the surface mass density azimuthal profile.

The maximum grain size shows an interesting, yet tentative profile, reaching a minimum of 1 mm at the beginning of the dust trap at $\theta$=218$^{\circ}$, and peaking at $\theta$=99$^{\circ}$ with a value of 4 mm. The mean maximum grain size along the dust trap is (1.5$\rm \pm$0.3) mm, compared to (2.2$\rm \pm$0.8) mm for the whole disk, and (2.6$\rm \pm$0.5)mm for the whole disk excluding the dust trap. Thus, there is a tentative decrease in the maximum grain size along the dust trap, with a size that is 1.7 times smaller than that outside the dust trap. Again, we note that the variations in grain size with azimuth are all within the 16-to-84\% percentiles, and thus the observed trends are subject to large uncertainties. For instance, at the position of the $a_{\rm max}$ minimum, the 84\% confidence level is 6 mm, and the 16\% confidence level is 0.2 mm. Adding more bands to our model comparison would help reduce the fitting uncertainty, and thus help us elucidate the actual value of the maximum grain size. Summarizing, and given the uncertainty associated with the maximum size, we conclude that our results are consistent with a grain size that is constant in azimuth along the ring, with a mean value of 0.22$\rm \pm$0.08 cm. This is overall consistent with the decaying vortex model proposed by \citet{Fuente2017}.

\subsection{Dust radiative transfer calculations}
\label{sec:RTcalcs}
In addition to the dust model presented in the previous section, we also carried out new dust radiative transfer calculations based on the hydrodynamical simulation in \citet{Fuente2017}, where a 2 Jupiter-mass planet at about 96 au was proposed to shape the mm cavity and the non-axisymmetric ring of emission outside it. In that simulation, a vortex forms at the outer edge of the planet gap due to the onset of the Rossby-Wave Instability, where dust gets efficiently trapped. The vortex ends up decaying due to the viscous turbulent evolution of the disc, and by doing so it progressively loses its dust which progressively acquires a near-axisymmetric distribution along the outer edge of the planet gap. This scenario was proposed by \citet{Fuente2017} to explain the unexpectedly smaller azimuthal contrast ratio of the ring emission at 2.2 mm than at 1.1 mm. A similar scenario was proposed for the non-axisymmetric ring of emission just outside the sub-mm cavity in the protoplanetary disc around MWC 758 in \citet{Baruteau2019}. We further stress that in \citet{Fuente2017} only the orbital radius of the putative planet responsible for the cavity and the ring emission could be predicted, not its position angle.

The new radiative transfer calculations presented here use the same parameters as in \citet{Fuente2017} except for (i) the dust composition, which is now assumed to be a mixture comprised of 70\% water ices and 30\% astro-silicates (to better reflect the fact that, at the radial distance of the ring, where the dust temperature is about 40K, one may expect dust grains to be coated with water ice mantles), (ii) the disc distance, which is now taken to be 162.9 pc \citep{GAIA2018}, and (iii)  the slope of the dust size distribution is changed from -3.5 to -3.0, which results in azimuthal contrast ratios at 1.1 and 2.2 mm in much better agreement with observations. Radiative transfer calculations were done with the public code \href{https://www.ita.uni-heidelberg.de/~dullemond/software/radmc-3d}{RADMC-3D} \citep{RADMC-3D-2012}. The post-processing of the Dusty FARGO-ADSG hydrodynamical simulation into input parameters for RADMC-3D was done with the public python code \href{https://github.com/charango/fargo2radmc3d}{fargo2radmc3d}; all details about the post-processing procedure can be found in \citet{Baruteau2019}. Note that the dust properties in the dust radiative transfer calculations differ from those inferred from the isothermal slab model of the previous section, in particular, the maximum dust size along the ring is taken to be 1 cm instead of $\sim0.15$ cm. This is not a problem since our radiative transfer calculations primarily aim at a qualitative interpretation of the observed maps, especially how the azimuthal contrast ratio of the ring emission varies with wavelength.

\begin{figure*}
\centering
\includegraphics[width=0.3\hsize]{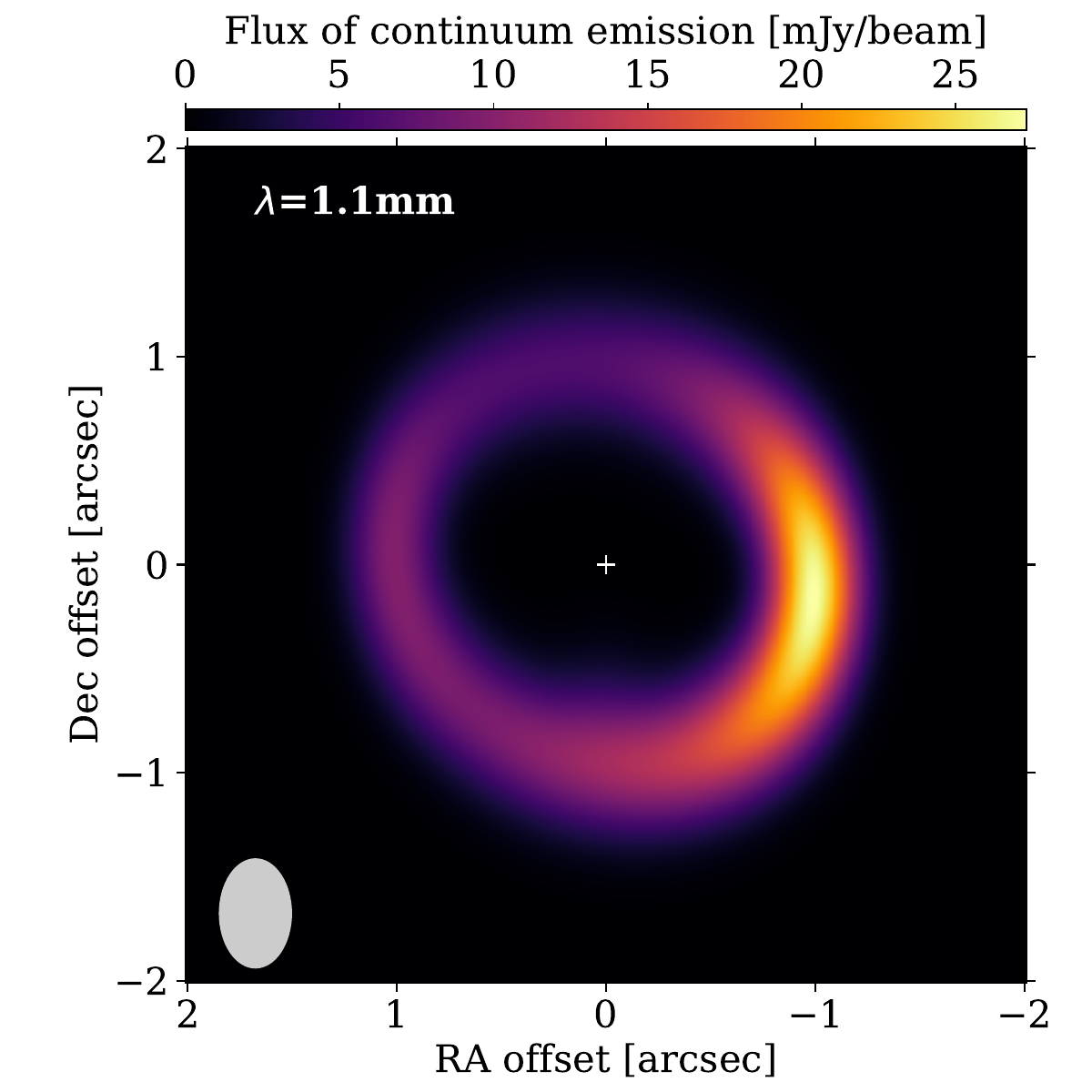}
\includegraphics[width=0.3\hsize]{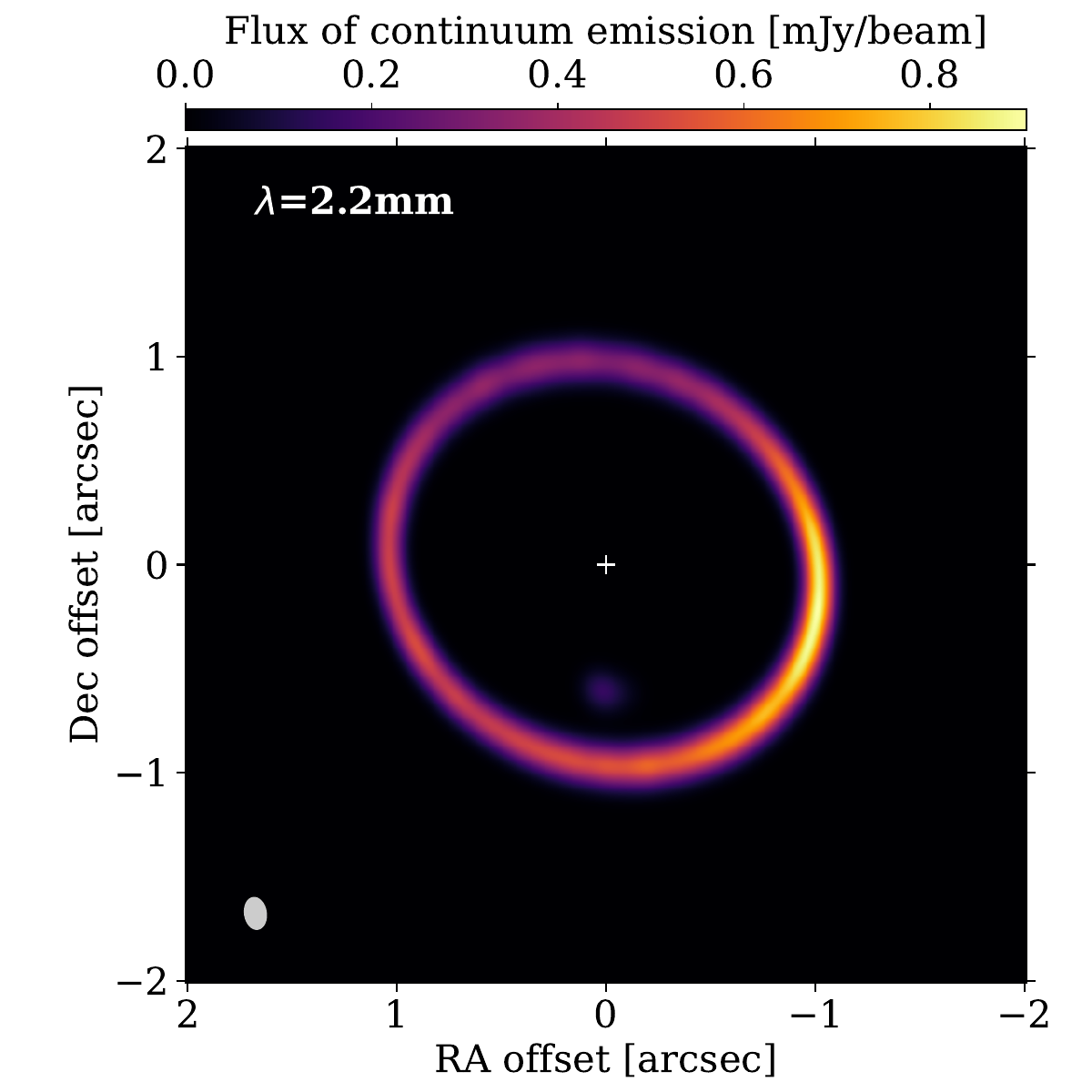}
\includegraphics[width=0.3\hsize]{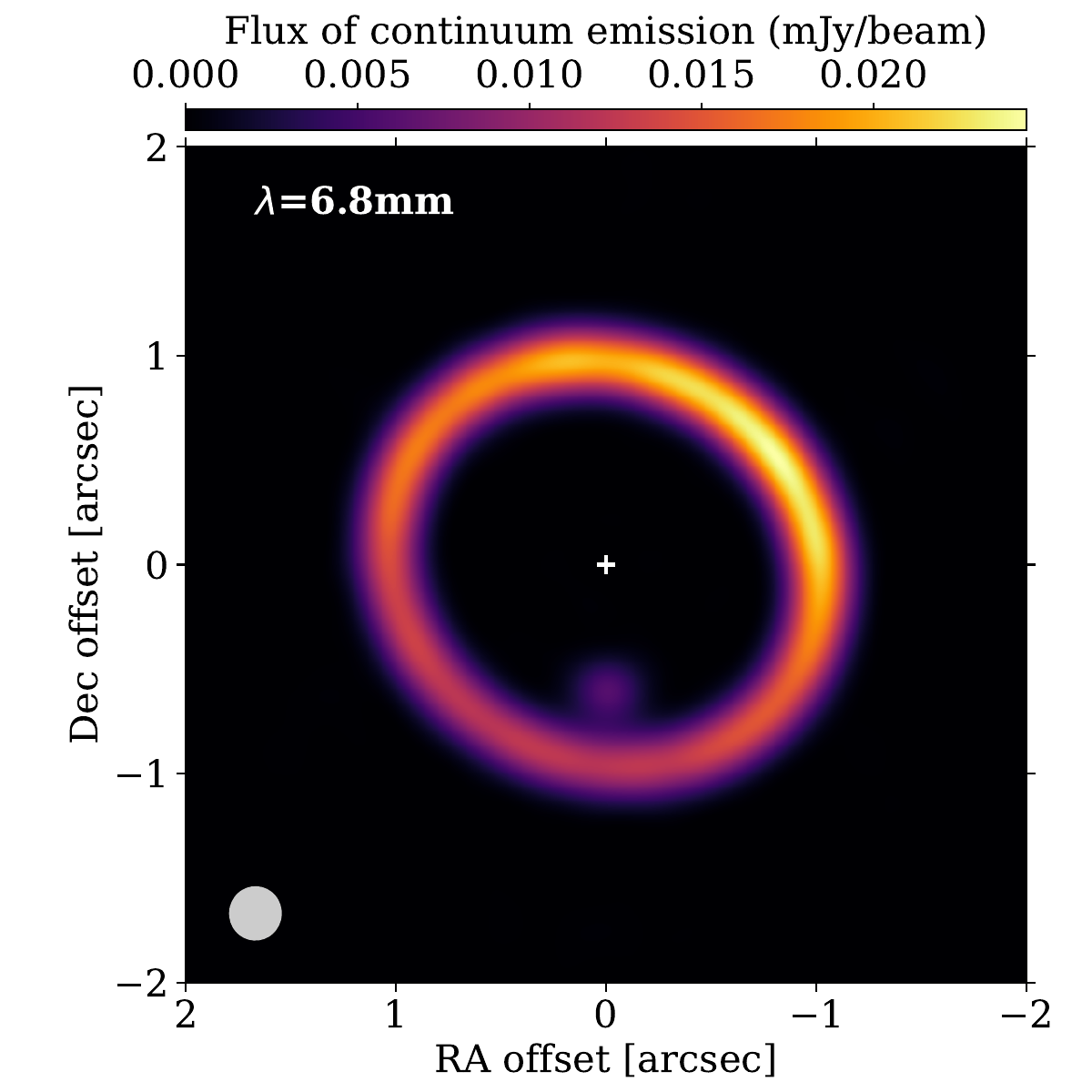}
\caption{\label{fig:model1v12v2}Synthetic maps of dust continuum emission at 1.1 mm (left), 2.2 mm (center), and 6.8 mm (right) with the same beams as the NOEMA and ALMA images shown in Fig.~\ref{Fig:cont_maps}, which are shown as grey ellipses in the lower left corner of each panel.}
\end{figure*}

\begin{figure}
\centering
\includegraphics[width=1.0\hsize]{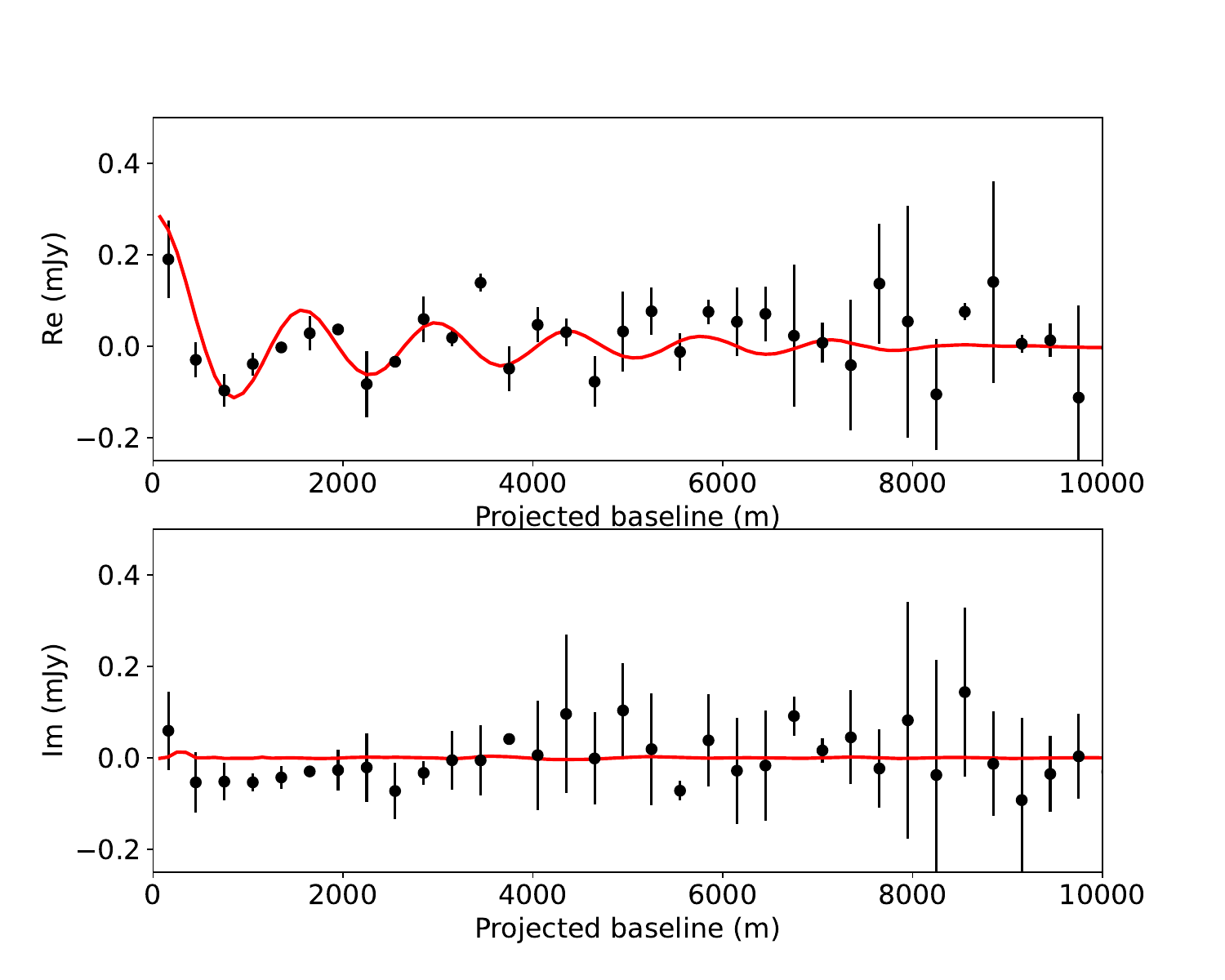}
\caption{\label{fig:model6v8_uvplane} Top: Azimuthal average of the real part of the visibilities of our RADMC-3D  model (red line, see Fig. \ref{fig:model1v12v2}) at 6.8 mm in the uv-plane. We also show our observational data rebinned by a factor of three at 6.8 mm without the central source for comparison. Bottom: same as the top panel but for the imaginary part.}
\end{figure}

Synthetic maps of continuum emission were computed at 1.1, 2.2, and 6.8 mm. They are displayed in Fig. 8. They show that the longer the wavelength, the larger the azimuthal extent of the continuum emission along the ring and the smaller its azimuthal contrast ratio (the same trend is found in our maps without beam convolution). In particular, the ring emission at 6.8 mm is nearly axisymmetric, in agreement with our scenario of a dust-losing decaying vortex. In Fig. \ref{fig:model6v8_uvplane}, we show the azimuthal average of the real and imaginary parts of our RADMC-3D model at 6.8 mm in the uv-plane compared to our observational data. The figure illustrates that, within the uncertainties, the model produces a reasonable fit, implying that the conclusions in Fuente et al. (2017) are compatible with our VLA observations at 6.8 mm.  Future, higher-sensitivity data are still needed to confirm this firmly. Note that if, on the contrary, the ring emission was related to an actively dust-trapping vortex, we would be in a situation similar to that in the disc around Oph IRS 48 and we should expect a compact VLA emission at position angles ~270$\rm ^{\circ}$ \citep{vanderMarel2015}. We note that there are discrepancies between the imaginary part of the model and the observed visibilities at short baselines. These differences suggest areas for improvement in the model's accuracy.

We now provide a more quantitative comparison for the ring emission between our synthetic and observed images. First, we see that the flux levels predicted in our synthetic images are in fairly good agreement with those in Fig.~\ref{Fig:cont_maps}, which can be appreciated by comparing the peak intensities along the ring: at 1.1 mm, the peak intensity is about 26 mJy/beam in the synthetic image and 8 mJy/beam in the NOEMA image; at 2.2 mm it is about 0.9 mJy/beam in the synthetic image and 0.3 mJy/beam in the ALMA image; at 6.8 mm it is about 24 $\mu$Jy/beam in the synthetic image and about 20 $\mu$Jy/beam in the VLA image. While the azimuthal contrast ratio does decrease from 1.1 to 2.2 mm in both synthetic and observed images, our synthetic values are larger: it is $\approx$4.3 at 1.1 mm (mean observed value: 3.1) and $\approx2.7$ at 2.2 mm (mean observed value: 2.3). Another difference is the radial extent of the ring emission at 2.2 mm which is smaller in our model than in the observation (by about a factor $\times2$). 

We finally stress that a faint, compact emission is slightly visible at position angles $\sim$180$^{\circ}$ to the north of the ring in our synthetic maps at 2.2 and 6.8 mm. It is due to mm-cm dust still around the L5 Lagrangian point located behind the planet in the azimuthal direction. Note that our model does not predict circumplanetary emission since in the hydrodynamical simulation Lagrangian dust particles entering the planet's Hill radius were automatically reinjected elsewhere in the computational grid to save computing time.

\section{Discussion}\label{Sect:discussion}
Our VLA observations of AB Aur at 7 mm are the longest wavelength-resolved observations of its circumstellar disk to date. Due to sensitivity limitations, we only detect emission at certain azimuths along the disks, as well as toward the compact central source. Using simple assumptions we have derived a total dust mass $\sim 0.09 M_{\rm Jup}$ (mean value), in good agreement with the value in \cite{Fuente2017}. \cite{Woitke2019} performed a detailed modeling of the SED and derived a dust mass $\sim 0.23 M_{\rm Jup}$, about 2.5 times larger, but still in good agreement given the uncertainties in the derivation. In \cite{Riviere2020} we derived a mean gas-to-dust ratio of 40, which results in a disk mass of $\sim 3.6 M_{\rm Jup}$. If we assume a gas-to-dust mass ratio of 100, our resulting mass for the disk around AB Aur is $9~M_{\rm Jup}$. 

This $\rm 9.0~M_{Jup}$ mass might be enough to form giant planets, depending on the planet formation efficiency. As pointed out by \cite{Andrews2005}, a fraction of the mass can be already stored in large grains and planetesimals that do not contribute to mm emission. Furthermore, there are indications that planets are already formed in AB Aur, such as the presence of a large cavity \citep{Pietu2005} and a dust trap \citep{Tang2012}. As recalled in the introduction, \cite{Fuente2017} proposed that the dust trap detected at mm wavelengths is due to a $\sim$2$\rm M_{Jup}$ forming planet at $\sim 96$ au from the star, a scenario that is further supported by the detection of a filamentary gas structure in \cite{Riviere2019}. Using the SUBARU Telescope and the Hubble Space Telescope, \cite{Currie2022} claimed the detection of a protoplanet at a projected distance of 93 au, very close to that predicted by \cite{Fuente2017}. The presence of a forming planet in AB Aur would be in line with present theories that predict planet formation could well start during the embedded phase of young stellar objects \citep{Segura-Cox2020}.

\begin{figure*}[t!]
\begin{center}
 \includegraphics[width=1.0\textwidth]{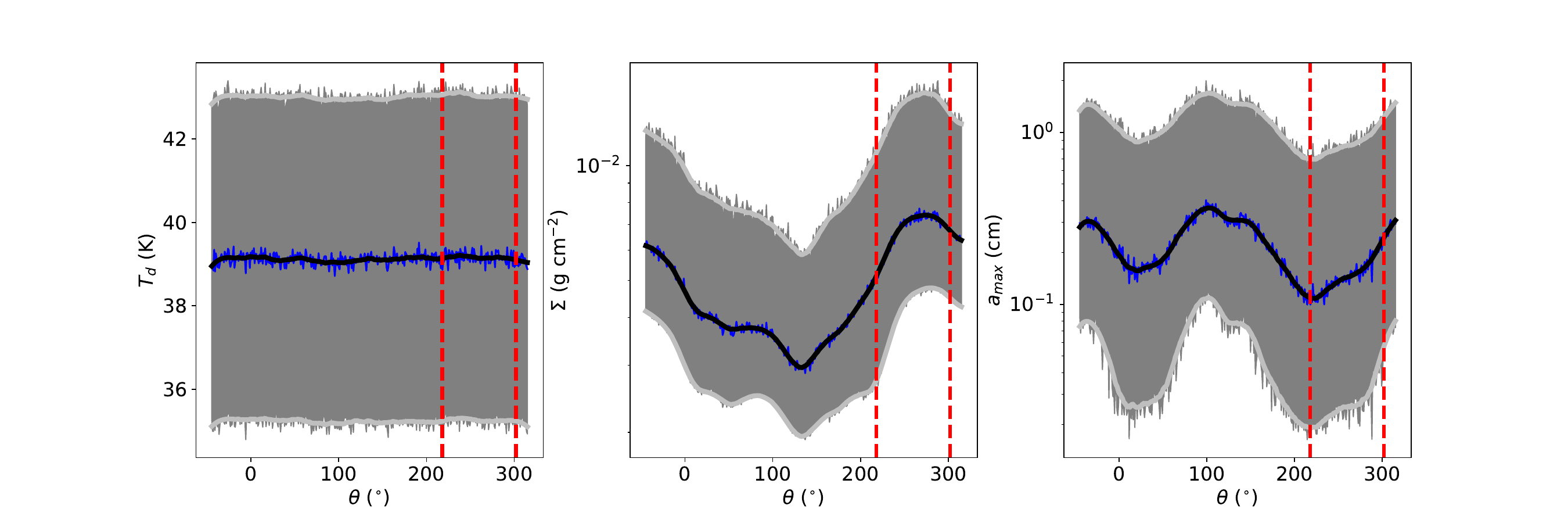}\\
  \caption{Results from fitting the azimuthal profiles of continuum emission with 1D isothermal slab models. The blue curve depicts the best value at each azimuth and for each parameter, while the thick black curve represents a smoothed version after applying a Savitzky-Golay filter with a window length of 51 and a polynomial of order 3. The dark grey shaded regions represent the 16\% and 84\% confidence levels. The light grey curves represent a smoothed version of the confidence levels after applying the same Savitzky-Golay filter. The red vertical dashed lines show the approximate azimuthal extent of the dust trap as traced by the 0.87, 1.1, and 2.2 mm continuum maps.}
 \label{Fig:dust_RT_model_results}
\end{center}
\end{figure*}

Our resolved observations of AB Aur allowed us to study its spectral energy distribution (SED) and to compute the spectral index at millimeter wavelengths. Using only resolved observations, we fitted the SED of the central source and the outer ring separately. In the outer ring, the spectral index is 2.3$\pm$0.2, in good agreement with the expected value in a cold protoplanetary disk with some degree of grain growth. The central source shows a spectral index of 1.1$\pm$0.3, which is inconsistent with pure dust thermal emission.  

\begin{figure}[t!]
\begin{center}
 \includegraphics[width=0.99\hsize]{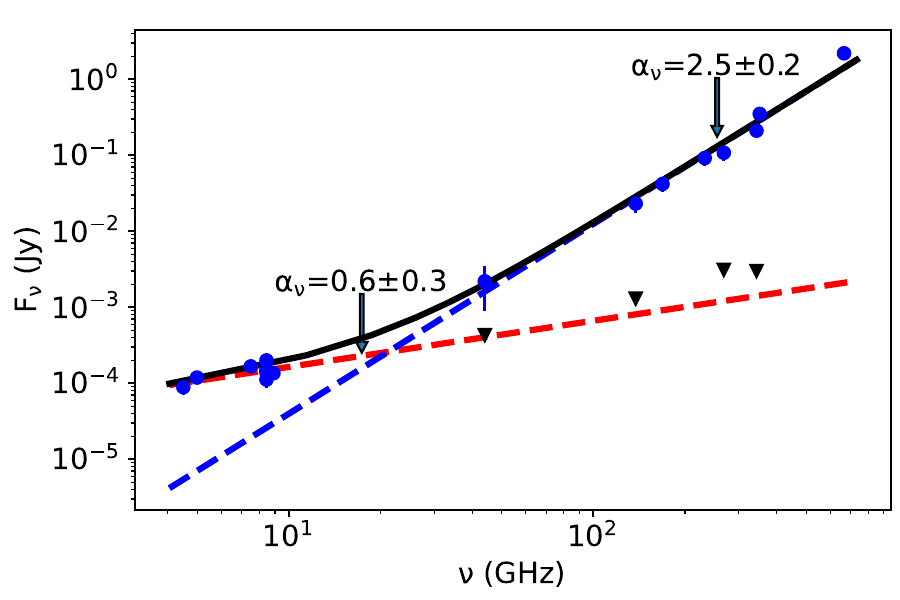}\\
 \includegraphics[width=0.99\hsize]{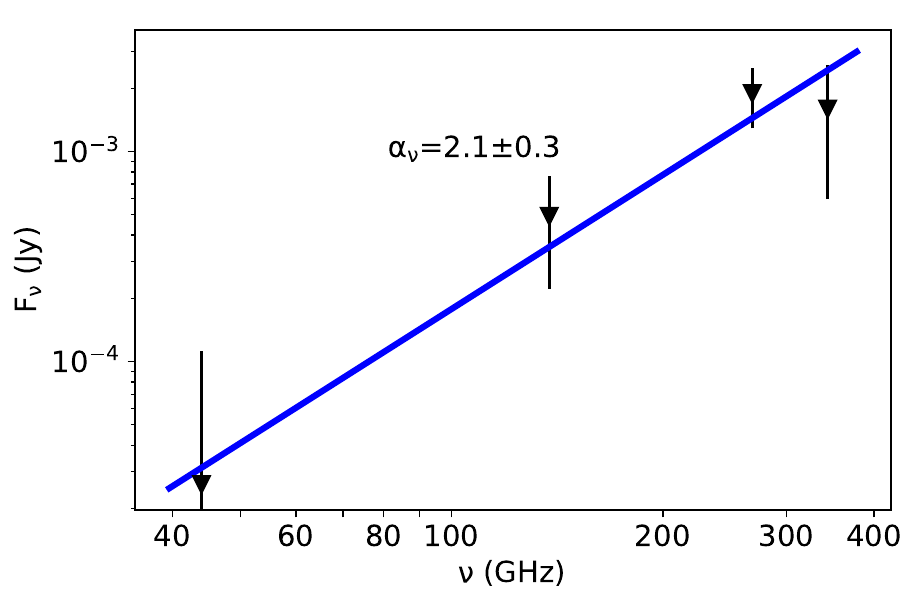}\\
  \caption{Top: spectral energy distribution from the sub-mm to cm wavelengths. The thick red line depicts a fit to cm data, while the blue one shows a fit to sub-mm and mm data. The thick black curve shows the combined fit.
   Blue dots are observations from this paper (mm data) and the literature (sub-mm and cm). Black triangles are fluxes from the central source alone. Bottom: spectral energy distribution at mm wavelengths for the compact source towards the center after subtracting the free-free contribution.}
 \label{Fig:SED_submm_to_cm}
\end{center}
\end{figure}

AB Aur has been extensively observed at centimeter wavelengths \citep{Guedel1989, Skinner1993, Rodriguez2007, Rodriguez2014}. The source has also been observed at sub-mm wavelengths using single-dish radio telescopes \citep{Sandell2011}. We have compiled the fluxes from all these observations covering sub-mm to cm wavelengths. The compiled data are summarized in Table \ref{Tab:SED}. With all these data we have built a SED aiming to derive the spectral index using the largest dataset possible. The result is shown in Fig. \ref{Fig:SED_submm_to_cm}. 

We fitted a single power law to the whole SED, but this resulted in a poor fit at most wavelengths. For this reason, we tried a second fit with different power-law exponents for the sub-mm and mm data on the one hand, and for cm data on the other hand (see thick black curve in Fig. \ref{Fig:SED_submm_to_cm}). This fit produces better results. The spectral index for the sub-mm and mm data (0.45 mm to 7 mm) is $\alpha_\nu$=2.5$\pm$0.1, while it goes down to $\alpha_\nu$=0.6$\pm$0.3 at cm wavelengths. This indicates that, while the sub-mm and mm data are dominated by dust thermal emission (i.e., the disk), the cm data is likely dominated by free-free emission. The VLA data at 7 mm is well reproduced with the same power law as the data at sub-mm and mm wavelengths. 

To illustrate this, we have included in the top panel in Fig. \ref{Fig:SED_submm_to_cm} the mm fluxes computed for the central source alone as black triangles. We can see that the mm fluxes from the central source are well reproduced by the fit to the cm data. Therefore, we conclude that the central source is dominated by free-free emission and that it contributes to most of the cm emission. Yet, the mm fluxes of the central source parts are higher than the fluxes extrapolated from the component dominating at cm wavelengths (i.e., $\alpha_\nu$=0.6). So we fitted a new power-law to the residuals obtained after subtracting the cm fit from the mm data for the inner disk. The resulting fit is shown in the lower panel in Fig. \ref{Fig:SED_submm_to_cm}. We obtain  $\alpha_\nu$=2.1$\rm \pm $0.3, which confirms that the mm emission observed towards the center is a combination of dust thermal emission and free-free emission. According to our fits, 93\% of the flux from the compact source in the center at 7 mm is due to free-free emission, and this fraction goes down to 31\% at 0.87 mm. Observations carried out with the Jansky Very Large Array (JVLA) revealed radio jet emission \citep{Rodriguez2014}. Our 2.2 and 6.8 mm observations of the central source are the mm counterpart of the cm emission detected by \cite{Rodriguez2014}. Now, our fit to the cm data confirms that the flux at 7 mm from the compact source in the center is dominated by free-free emission, as concluded by \cite{Rodriguez2014}. Regarding the dust thermal emission arising from the central compact source, the higher spectral index compared to the outer ring indicates that the dust in this inner region is dominated by smaller grains. 

The radial profiles of our $\alpha _{0.87{\rm mm}-2{\rm mm}}$ and $\alpha _{1{\rm mm}-2{\rm mm}}$  spectral index maps (Fig. \ref{Fig:sp_index_maps}) show a bump around 0.5$\arcsec$ that can be attributed to the presence of sub-mm sized grains inside the cavity. Such a bump is also observed in the map computed using the three bands (0.87, 1.1, and 2.2 mm), shown in panel c) in Fig.~\ref{Fig:sp_index_map_all}. This can be the consequence of dust filtration at the outer edge of a planet gap with mm-sized dust grains being trapped around the pressure maximum formed at the gap's outer edge while smaller grains better coupled to the gas keep on drifting inward of the gap \citep{Zhu2012}. Note that the dust filtration ability of a planet is also sensitive to the disk's aspect ratio and turbulent viscosity around the planet \citep{Ataiee18}. In the dust filtration scenario, dust gaps are expected to be deeper and wider with increasing grain size \citep{Fouchet2010} with the consequence that (sub)-mm cavities observed in the radio should be larger and deeper with increasing wavelength \citep{Zhu2012}.

Beyond the bump near 0.5$\arcsec$, spectral indices remain flat, with values in the range [2.2-2.8]. This is in contrast with observations of other protoplanetary disks, such as HD 163296 and TW Hya, where spectral indices grow with radius at all radii \citep{Dent2019,Macias2021}. In these two systems, dust rings and gaps are present, but no large cavities, while a large cavity ($\rm \sim$)100 au is observed in AB Aur. The cavity in AB Aur can be due to the presence of a forming planet, such as the one proposed by \cite{Currie2022}, \cite{Zhou2022}, and \cite{Fuente2017}. This scenario is further supported by the dust-filtering effects described in the previous paragraph. Further research is needed to understand the origin of this difference in the spectral index radial profile. 

We also studied the azimuthal distribution of the spectral index, but our results are not conclusive in this regard. Since the 7 mm map lacks emission in most of the azimuths, we rely on the study of the 0.87 mm to 1.1mm, 0.87 mm to 2.2 mm, and 1.1 mm to 2.2 mm spectral index maps. The azimuthal profile of the 1.1 to 2.2 mm spectral index map shows a decay of the spectral index in the dust trap region (see Fig. \ref{Fig:az_profile}). The decrease in the spectral index at the position of the dust trap indicates an enhanced population of large grains compared to smaller grains. \cite{Fuente2017} proposed that the disk around AB Aur has been sculpted by a 2$\rm M_{Jup}$ planet at a distance of 96 au. The interaction of such a planet with the disk resulted in the formation of a vortex. The authors showed that the azimuthal contrast ratio was larger at 1.1 mm than at 2.2 mm, indicating that the largest grains were less concentrated, a fact that they interpreted as an indication that the vortex originating the dust trap has started to decay, a result further supported by the present study.

\section{Summary and conclusions}\label{Sect:summary}
We have used ALMA, VLA, and NOEMA observations to study the dust properties of the protoplanetary disk around AB Aur. This study is part of a long-term study of the circumstellar disk around AB Aur \citep{Fuente2010, Pacheco2015, Pacheco2016, Fuente2017, Riviere2019, Riviere2020, Riviere2022}. The main goal of this work was to study the spectral index spatial distribution and its relation with continuum and molecular emission. In the following, we summarize our main results

1. We have resolved the continuum emission at 6.8 mm towards AB Aur for the first time. Our map shows a bright compact source towards the center surrounded by knots of emission at a distance of $\sim$1$\arcsec$. The total flux at 6.8 mm is 2.2$\rm \pm$1.1 mJy. We also computed continuum fluxes of 92$\rm \pm$19 mJy at 1.1 and of 23$\rm \pm$6 mJy at 2.2 mm.

2. We derived radial profiles of the three maps. The maps at 2.2 and 6.8 mm peak towards the center, while the low-angular resolution map at 1.12 mm peaks at 0.9$\arcsec$ from the center. The maps at 2.2 and 6.8 mm also show a secondary peak at 1$\arcsec$ from the center. 

3. We computed the SED from submm to cm wavelengths using our new observations and archival data from the literature. We derived a spectral index of 1.0$\rm \pm$0.2 toward the central source, and 2.3$\rm \pm$0.7 in the outer disk. We also collected (sub)mm and cm data from the literature and built a SED covering three orders of magnitude. The cm data is best reproduced by a slope $\alpha$=0.6$\rm \pm$0.3 compatible with free-free emission, while (sub)mm data is best reproduced by a slope $\alpha$=2.5$\rm \pm$0.1. By subtracting the free-free contribution to resolved emission in the (sub)mm we managed to retrieve a spectral index of 2.1$\rm \pm$0.3, indicating that a population of large grains is also contributing to the observed emission.

4. We computed spectral index maps, as well as their radial and azimuthal profiles. The radial profiles are flat at the position of the outer disk, indicating no radial variation of the spectral indices, other than a subtle decay in the outermost regions. We observed a bump in radial profiles around 0.5$\arcsec$ that we tentatively attributed to the presence of sub-mm-sized grains inside the cavity.

5. A 1D isothermal slab model for the dust emission along the ring outside the millimeter cavity indicates no modulation of the maximum grain size with azimuth within the statistical uncertainties, with a mean maximum grain size of 0.22$\rm \pm$0.08 cm. 

6. The fact our VLA image shows knots of $5\sigma$ emission all along the dust ring and not only near the position angle of the peak intensity at 1.1 and 2.2 mm suggests that the 7 mm continuum emission along the ring has a low azimuthal contrast ratio. This is consistent with the scenario proposed in \citet{Fuente2017} of a dust-losing decaying vortex at the outer edge of a planet gap to explain the decrease in the azimuthal contrast ratio of the ring emission with increasing millimeter wavelength.

\begin{acknowledgements}
P.R.M. and A.F. thank the Spanish MICIN for funding support from PID2019-106235GB-I00. We thank the ALMA Archive and Data Retrieval (EU) team for providing us with calibrated data for the 0.87 mm ALMA observations. A.F. acknowledges funding from the European Research Council advanced grant SUL4LIFE (GAP-101096293). A.R. has been supported by the UK Science and Technology Research Council (STFC) via the consolidated grant ST/W000997/1 and by the European Union's Horizon 2020 research and innovation program under the Marie Sklodowska-Curie grant agreement No. 823823 (RISE DUSTBUSTERS project). DNA acknowledges funding support from Fundaci\'on Ram\'on Areces through its international postdoc program. G.A. and M.O. acknowledge financial support from grants PID2020-114461GB-I00 and CEX2021-001131-S, funded by MCIN/AEI/10.13039/501100011033. We acknowledge the community effort devoted to the development of the following open-source packages that were used in this work: \texttt{numpy} \citep[\texttt{numpy.org},][]{numpy20}, 
\texttt{matplotlib} \citep[\texttt{matplotlib.org},][]{matplotlib}, \texttt{SciPy} \citep[\texttt{scipy.org},][]{2020SciPy}, and \texttt{astropy} \citep[\texttt{astropy.org},][]{astropy:2013, astropy:2018, astropy:2022}.

\end{acknowledgements}

 \bibliographystyle{aa} 
\bibliography{biblio}

\end{document}